# Biochars at the molecular level.

# Part 1 – Insights into the molecular structures within biochars.

**Rosie Wood[a], Ondřej Mašek[b], Valentina Erastova[a]***

[a] School of Chemistry, University of Edinburgh, Joseph Black Building, David Brewster Road, King's Buildings, Edinburgh, EH9 3FJ, United Kingdom

[b] UK Biochar Research Centre, School of GeoSciences, University of Edinburgh, Crew Building, Alexander Crum Brown Road, King's Buildings, Edinburgh, EH9 3FF, United Kingdom

*valentina.erastova@ed.ac.uk

**Abstract**

Biochars are black carbonaceous solids produced through biomass pyrolysis under conditions of little or no oxygen. Whilst their properties are well studied, and their applications numerous, the underlying molecular structures within biochars still need to be defined. This raises a substantial barrier to the molecular modelling of biochars and has limited computational study of these materials, despite the advantages such techniques. In Part 1 of the work, we critically assess the analytical techniques used to characterise biochars and use this information to gain atomic-resolution insights into biochars' molecular compositions and nanostructures. We compile experimental data from public domains and use it to inform us of the trends in biochar properties. Through this work, we create a holistic understanding of biochar's chemical, physical and molecular properties and lay the foundation for Part 2 – the development of realistic molecular models of biochar materials for molecular dynamics simulations.

**Keywords**

Biochar, charcoal, molecular structure, molecular model, analytical techniques, database

# 1. Introduction

*Biochars*, *charcoals* and *chars*, are black carbonaceous solids produced through pyrolysis of biomass, an organic material from previously living organisms as well as anthropogenic waste materials, under conditions of little or no oxygen. The properties of resulting materials are incredibly variable and heavily influenced both by the starting feedstock, and the conditions used during pyrolysis. This variability renders biochars suitable for diverse applications. Water decontamination, carbon capture and soil-nutrient enhancement are but a few examples.[1]

A clear understanding of structure-property relationships within biochars is vital when attempting to identify optimal feedstocks and production conditions for a given application. This can be achieved through a series of systematic and well-thought-out experiments. However, this process is highly laborious, resource-intensive, and due to the heterogeneity of biochars, must be repeated many times to reduce uncertainties. Computational techniques, such as molecular dynamics (MD), have been gaining popularity due to their ability to describe system's structure-property relationships. Modelling allows for a systematic and reproduceable investigations, testing hypothesis and informing experiment. However, whilst the properties of biochars are well understood individually,[2] relatively little is known about the underlying molecular structures. As a result, realistic molecular models of biochars are virtually non-existent, drastically limiting the scope of MD study of these materials.

The goal of this work is to gain insights into the molecular structures of biochars, understand how these structures vary with production conditions, and to create and distribute a set of biochar molecular models to be used in MD studies. This work is in two parts: Part 1 – Insights into the molecular structures within biochars (current article) and Part 2 – Development of realistic molecular models of biochars (see Ref. 3). We begin by setting biochars in context and with a brief discussion of feedstocks and their influence on biochar properties. We then examine the different analytical techniques used to characterise these materials along with their ambiguities, errors and uncertainties. This allows us to assess the information gained by each technique critically. We focus predominantly on techniques which yield molecular-level insights into biochars, thereby adding to our understanding of biochars' molecular make-up. We also collect experimental data from the literature into an open database (accessible via https://github.com/Erastova-group/Biochar_MolecularModels). And finally, we use this data to gather quantitive insights into the influence of one of the most significant production parameters – the highest treatment temperature (HTT) – on the properties of biochars. We use this data to illustrate the relationships between HTT and biochar properties and, therefore, to infer the influence of HTT on molecular structure.

The collected data, supported by the deeper understanding of biochars' molecular make up, aids us in the development of a set of realistic molecular models of biochars, presented in the Part 2 of this work.[3]

## 1.1. Natural and manufactured carbonaceous materials

Biochars are formed through the carbonisation of organic matter during pyrolysis. As they share similar precursors (lignocellulosic biomass), it is useful to consider biochars alongside natural carbonaceous solids, such as *coals* and *kerogens*. These may be regarded as natural analogues to biochars,[4] and, as a result, many of the analytical techniques used to characterise biochars originate from the characterisation of these natural materials.

Coals are defined as carbonaceous sedimentary rocks containing at least 50% wt. organic matter. They are usually formed from terrestrial biomass, such as woody organic matter.[4,5] On the other hand, kerogens are carbonaceous solids derived from sedimentary rocks containing less than 50% wt. organic matter. They are defined as being insoluble in non-oxidising acids, bases and organic solvents.[4] There are three main types of kerogens: types I and II, which originate from marine biomass, and type III, which originates from terrestrial biomass.

The relationship between biomass, biochars, coals and kerogens are displayed can be presented as ratios of key organic elements H:C vs O:C with a *Van Krevelen diagram*, shown schematically on **Figure 1**. Dirk Van Krevelen originally developed these diagrams to assess and compare the compositions of coal samples.[6,7] Biomass, a collective term used to describe a wide range of plant-derived materials, occupies a large area at the top right corner of the diagram in **Figure 1**, indicating the presence of relatively large amounts of oxygen and hydrogen in its structure. On the other hand, the three materials derived from biomass – biochars, coals and kerogens – occupy more discrete bands, indicating their distinct properties and chemical compositions.[4]

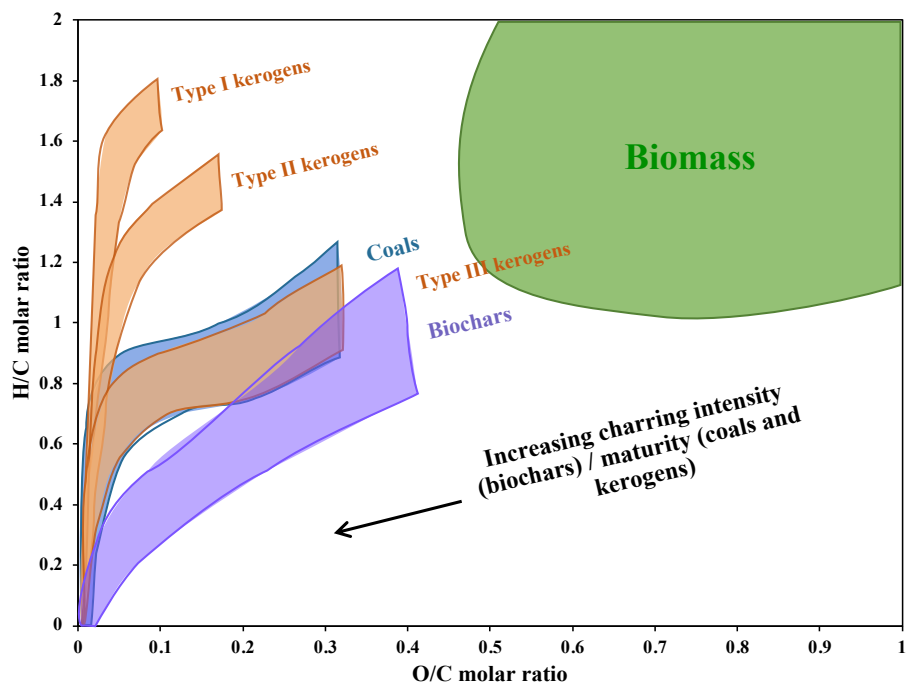

**Figure 1.** A schematic guide to the relationship between carbonaceous materials (biomass, biochars, coals and kerogens), shown as a Van Krevelen diagram, data taken from Refs 8–16.

Biochars, coals and kerogens originate from biomass and begin their existence with chemical compositions not vastly dissimilar from it. However, with increasing processing (for biochars) or maturity (for coals and kerogens), hydrogen and oxygen are lost from these materials, leading to a relative increase in carbon content and a decrease in H/C and O/C ratios.

Whilst the chemical compositions of coals and type III kerogens almost entirely overlap – a consequence of their similar origins (lignocellulosic biomass) – the compositions of biochars are distinct. This divergence is evidence of the differences between the structures created during pyrolysis (higher temperature, shorter times) and the natural maturation (higher pressure and geological timescale) of biomass.[4,17] Nevertheless, biochars possess enough similarity to coals and kerogens to allow the analytical techniques used in characterising those materials to be transferred into their study.

### 1.2. Feedstocks and plant biomass

Any form of dried biomass can be used to produce a biochar. Plant biomass, commonly derived from agricultural waste, such as sawdust, wood chips, straw or husks, is a notable example. It can be thought of as consisting of two fractions – an organic fraction, containing carbon-rich molecular structures, and an inorganic fraction, containing a range of mineral compounds and water. An example schematic composition of dried plant biomass is shown in **Figure 2**.

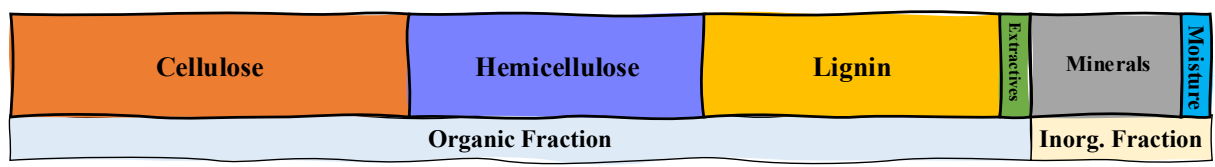

**Figure 1. A schematic representation of a macromolecular composition of plant biomass after drying.**

The organic fraction of plant biomass consists primarily of three types of biopolymers: *cellulose*, *hemicellulose* and *lignin*. Upon pyrolysis, these biopolymers thermally decompose in characteristic manners to produce biochar, bio-oil and syngas. Cellulose and hemicellulose have low thermal stabilities and decompose between temperatures of approximately 300-400ºC and 200-300ºC, respectively.[18–20] During their decomposition, a large proportion of their masses are lost as bio-oils and syngas and, as a result, cellulose and hemicellulose contribute only moderately to biochar yields.[18,19,21] Lignin polymers, on the other hand, have higher thermal stability. Their thermal decomposition occurs across a wide temperature range due to their heterogeneity, beginning at approximately 200ºC and continuing to temperatures as high as 900ºC.[18,19] Lignin decompose with relatively low mass loss and are, therefore, the main contributors to biochar yields.[18–22]

The relative quantities of cellulose, hemicellulose and lignin within a feedstock heavily influence the properties of the biochar it produces.[19,21,23,24] Lignin polymers have the highest carbon content and often considered the most influential of all biopolymers. Due to lignin's carbon content and stability, lignin-rich feedstocks generally found to produce highly-carbonised biochars.[25]

The mineral constituents of biomass are mostly non-volatile and are primarily converted to ash during pyrolysis. As a result, feedstocks with high mineral content produce biochars with high ash contents, while those with low mineral content produce biochars with low ash contents.[12,13,26–28] These inorganic minerals are also known to have both catalytic effects, which can promote a variety of reactions during the carbonisation process, and stabilising effects, which may protect or inhibit the decomposition of certain molecular structures.[1,12,23,26,29–31] As a result, high ash biochars may contain distinctly different molecular structures than their low ash counterparts.[12,32]

## 2. Characterisation of biochars

In the following section, we discuss the characterisation of biochars, focusing on the different methods used and the molecular descriptors obtained by each. Owing to the variety of techniques available, many properties of biochars can be measured. However, it is essential to understand the limitations and sources of error within these measurements, as well as how different techniques can complement each other to obtain a holistic understanding of the properties of biochars.

### 2.1. Chemical composition

The chemical composition of a biochar can be determined through *proximate* and *ultimate* analyses. The proximate analysis allows for the identification of fractions of moisture, fixed carbon, volatile matter and ash in a material. In contrast, the ultimate analysis provides insights into the discrete elements within the material. **Figure 3** shows a schematic representation of biochars composition that can be obtained through these methods respectively.

#### 2.1.1. Proximate composition

As with biomass, biochars can be thought of as being composed of organic and inorganic fractions, as shown in **Figure 3**.[14] The organic fraction consists of heterogenous organic matter with varying thermal stabilities: components which are stable at high temperatures are known as *fixed carbon* (FC), whilst those which decompose at high temperatures are known as *volatile matter* (VM).[5,33] At room temperature, however, both FC and VM exist as solids. The inorganic fraction of a biochar consists of ash and a small amount of residual moisture.

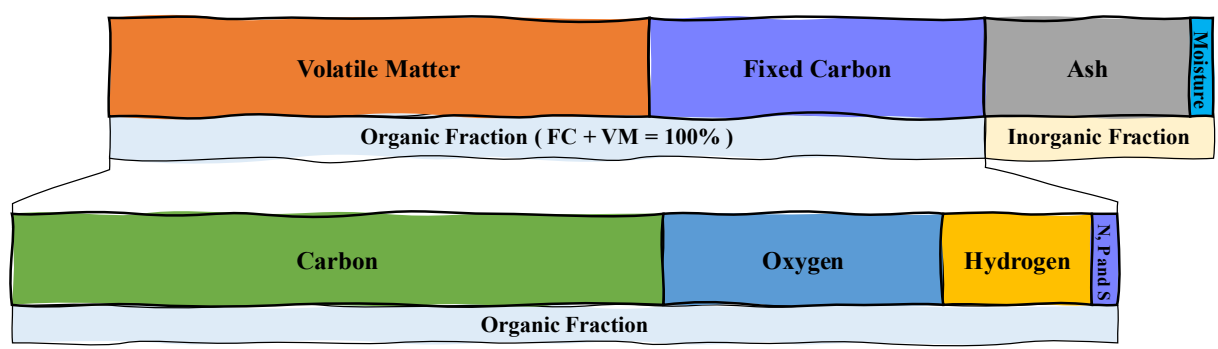

Figure 3. A schematic guide to the chemical composition of biochars. The upper band depicts the composition of a material, including both organic (volatile matter and fixed carbon) and inorganic (ash and mosture) fractions; the lower band depicts the elemental composition of the organic fraction only.

Proximate analysis has its roots in fuel chemistry, where it is used to determine the fractions of moisture, ash, VM and FC in a material at operational temperature (approximately 1000ºC).[5] It has been translated into the study of biochars and is now one of the most commonly used characterisation techniques within biochar research.[16,23,26,29,34–36]

Traditionally, proximate analyses are carried out using the standard procedures detailed in ASTM D1762-84 – "Standard Test Method for Chemical Analysis of Wood Charcoal".[33] In brief, moisture is defined as the mass lost when a sample is heated to 105ºC in an inert atmosphere, VM is defined as the mass lost (minus that due to moisture) when a sample is heated to 950ºC in an inert atmosphere, and ash is defined as the mass remaining after combustion of a sample at 750ºC. FC is then calculated by difference. Uncertainties in these measured values primarily arise from unwanted or unaccounted-for mass changes during heating/combustion (e.g. adsorbed atmospheric moisture, partial combustion, volatilisation of ash compounds), fluctuations in the temperature of the furnace and sample inhomogeneity.[5,36,37] Averaging across multiple measurements helps to reduce the error.

More modern techniques, based upon thermogravimetric analysis (TGA), are also used and measure mass loss in situ throughout the heating process.[16,27,36,38,39] TGA-based proximate analyses can be automated through use of a pre-set heating programme, making it more convenient and reliable than the traditional method. However, TGA is still susceptible to many of the same issues outlined above and, since very small sample sizes are often used, TGA-based analyses can be especially affected by sample inhomogeneity.[36] Averaging across multiple samples is, therefore, essential when using TGA.

Although useful for determining the proportions of inorganic to organic matter within biochars, proximate analyses give no absolute insights into their molecular structures. These molecular structures must, therefore, be characterised using other techniques. Nevertheless, knowing the ash and moisture content of a sample is critical when carrying out further characterisation as, often, both ash and moisture can strongly interfere with results.[23,26,36,40] In many cases, samples may be dried and de-ashed before any further characterisation to limit these effects, particularly for high ash biochars.

### 2.1.2. Ultimate composition

The organic fraction of a biochar consists predominantly of carbon, hydrogen, oxygen and, to lesser extents, nitrogen, phosphorus and sulphur (**Figure 3**). The relative amounts of these elements within a biochar sample are measured by ultimate analysis.

Ultimate analyses rely on the characterisation of the gases produced after the complete combustion of a dried sample of known mass in an oxygen-rich atmosphere. The gases produced are collected, separated, and the quantities of $CO_2$, $H_2O$, $N_2$ and $SO_2$ are determined. These are then used to calculate the organic C, H, N and S content, respectively, of the original sample. While O content can, in principle, be determined by CNHS analyser, in practice this is often done based on the assumption that the quantities of other elements in the sample are negligible, so the amount of oxygen is calculated as a difference.[36]

To gain insights into the molecular structures within biochars, the elemental composition determined by ultimate analysis must only correspond to the organic fraction of a sample. However, biochar ash often contains carbon in the form of carbonates,[41] resulting in a measured C content corresponding to both the total organic and total inorganic carbon (TOC and TIC, respectively) fractions.[36,37] To remove the influence of these carbonates, samples should be washed before analysis, yet in many cases, this is not carried out.[10–12,29,42–47] When the ash content of a sample is low, this is passable, as the contribution of TIC to the total carbon (TC) content is likely negligible, and so TOC can be approximated as TC.[10,26] However, when the ash content of a sample is high, TC is not a good indicator of TOC and should not be used as such.[26] Knowledge of sample's ash content via proximate analysis is, therefore, crucial when examining the results of an ultimate analysis. Furthermore, in highly carbonised biochars with significant amounts of combustion-resistant carbon, incomplete combustion can also be problematic and lead to an underestimation of C contents. Utilising a catalyst to encourage combustion can help reduce this error.[36]

Within biochar research, the results of ultimate analyses are often used to calculate molar H/C and O/C ratios.[10,12,15,16,22,48–52] These ratios can be used to assess the extent of carbonisation of a sample – with lower H/C and O/C ratios indicating a higher degree of carbonisation – and are often presented against each other in a Van Krevelen diagram, such as that shown in **Figure 1**.[8–16] Strictly speaking, these ratios should be calculated using only the TOC content to give clear insights into the carbonisation of the organic matter within biochars.[5] However, in cases where the organic C contents has not been explicitly measured, these ratios are usually calculated using the TC contents.[11,12,43–46] As a result, many the reported H/C and O/C ratios of high ash biochars may be misleading.[10]

Ultimate analyses provide direct insights into the elemental compositions of biochars and are, therefore, incredibly useful when attempting to understand their molecular structures. However, due to the multitude of ways in which atoms can be arranged whilst retaining the same elemental composition,[53] ultimate analyses alone cannot be used to predict the molecular structures within a sample.

## 2.2. Aromatic molecular structures

The aromatic molecular structures within biochars can be defined by their *aromaticity indices* (the proportion of aromatic carbons) and their *aromatic domain sizes* (the number of conjugated aromatic rings within polyaromatic domains).[34,50,54–60] The schematic in **Figure 4** shows the relationship between these two properties. Three simple compounds which can be used as illustrative examples of these properties are benzene, naphthalene and butylbenzene. Both benzene and naphthalene have aromaticity indices of 100%, whereas butylbenzene, with its alkyl group, has an aromaticity index of 60% (6 out of 10 carbons being involved in aromatic molecular structures). On the other hand, both benzene and butylbenzene have aromatic domain sizes of one ring, whereas naphthalene has an aromatic domain size of two rings. This exemplifies the importance of both properties when attempting to define the aromatic molecular structures within a biochar.

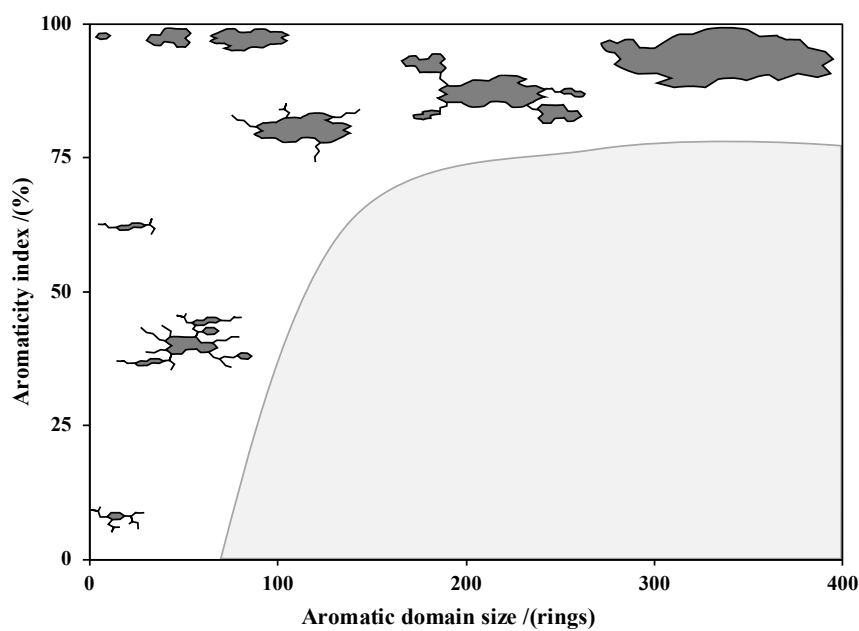

Figure 4. Schematic guide to the *aromaticity indices* and *aromatic domain sizes* of biochars.

$^{13}$C Solid-State Nuclear Magnetic Resonance (ssNMR) can detect the various chemical environments within a biochar sample. It is then possible to calculate its aromaticity index using the relative proportion of aromatic and non-aromatic chemical environments.[34,56,59,60] Highly conjugated aromatic chemical environments can, however, be challenging to observe/detect using a typical $^{13}$C ssNMR set up[34,58,60,61] and lead to error when calculating aromaticity indices of highly carbonised biochars. Alternative set ups can be used to reduce this effect,[58,60] yet complex spectra and difficulties in deconvoluting peaks may still yield relatively high uncertainties when calculating aromaticity indices.

Aromatic domain sizes are even more difficult to characterise, though a number of techniques have been developed.[34,50,63,52,56–62] The simplest method uses H/C to estimate the average aromatic domain size, $n$, of a biochar.[52,59] This is done by solving **Equation 1** for $n$, following the assumption that all organic carbon within the sample is in the form of 6-membered aromatic rings solely consisting of C and H and that aromatic domains are distributed as a 2-dimensional grid of $\sqrt{n}$ by $\sqrt{n}$ aromatic rings, as shown in **Figure 5a**.[52] Aromaticity indices are then assumed to be 100%.

$$\frac{H}{C} = \frac{2\sqrt{n} + 1}{n + 2\sqrt{n}} \qquad \textbf{(Equation 1)}$$

Generally, the aromatic domain sizes predicted using this method are expected to be underestimated when the actual H/C is higher than the idealised one (from the **Equation 1**) for the same domain size. The H/C ratio will be higher when aromatic domains are distributed irregularly, yielding a more significant number of aromatic edge sites (as shown in **Figure 5b**) and/or when functional groups with higher H/C than defined in the **Equation 1** are present (as shown in **Figure 5c-d**). Only in a small number of cases, the functional groups or carbon atom substitutions will reduce the overall H/C ratio (as shown in **Figure 5e**), then the aromatic domain sizes will be overestimated. Despite these apparent downfalls, this method offers a quick and easy route to estimating aromatic domain sizes within a sample. However, these values should generally be taken only as rough guides.

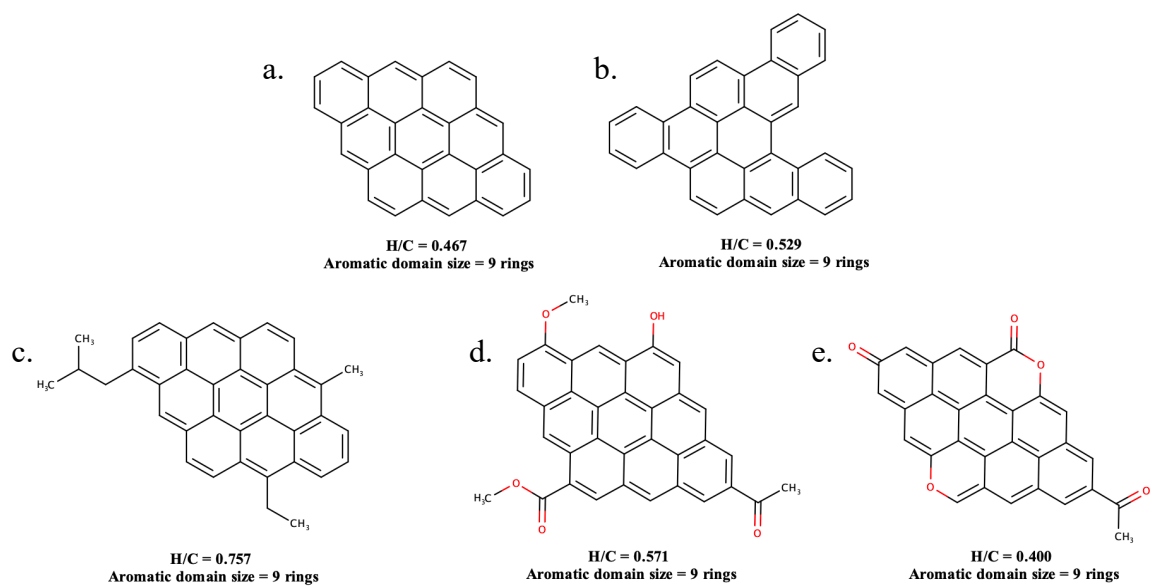

**Figure 5. Variation in H/C ratio for structures with the aromatic domain size of nine rings.**

High-resolution Mass Spectrometry (MS) is a collection of techniques, all based upon separation of ionised particles by their time of flight in the electromagnetic field. MS instriument measures the mass-to-charge ratio, that can be matched to the known molecular fragments, for example spacings of 14 Da indicate $N^+$, as well as $CO^{2+}$ and $CH_2^+$ and spacings 72 Da indicate $C_6$ (i.e. aromatic moieties).[29,64–66] Furthermore, the separation provide insights into the shape and interaction of the domains with the electromagnetic field. There are many types of MS, depending on the methods of ionisation and detection. There are two major techniques of ionisation – electrospray ionisation (ESI), that first requires the sample to be in a solution, and laser-desorption ionisation (LD-MS, e.g. MALDI), applies rapid heat vaporising the sample. Therefore, when it comes to the solid amorphous material, such as biochar, the main limitation of MS is in the accessibility of the structures for ionisation, i.e. the obtained results are unlikely to be representative of the overall material.[23,47,66–68] However, the information obtained will still yield interesting and useful insights into the molecular structures within a biochar sample and, especially, their aromatic domain sizes.

Aromaticity indices and aromatic domain sizes must both be known to build a complete understanding of the aromatic molecular structures within biochars. However, more information is still needed to understand these molecular structures' morphologies and how they pack together into nanostructures within the overall material.

## 2.3. Graphitisation and crystallinity

Carbon-rich materials can be thought of as either *graphitising* or *non-graphitising*, as shown in **Figure 6**.[54,69] Graphitising materials are those which form nanostructures resembling crystalline graphite when treated at temperatures of 2200 – 3000ºC, whilst non-graphitising materials are those which resist this ordering process and remain largely amorphous even when heated to temperatures exceeding 3000ºC.[54,70] The structural differences between these two types of material are believed to result from the formation of defects, such as non-hexagonal rings, within their aromatic nanostructures.[66,70,79,80,71–78] These defects introduce curvature and act to disrupt packing and limit long-range ordering. In general, biochars possess only minimal graphitic crystallinity and, though rarely heated to temperatures as high as 2200ºC, are thought to be non-graphitising.[8,11,55,81–84]

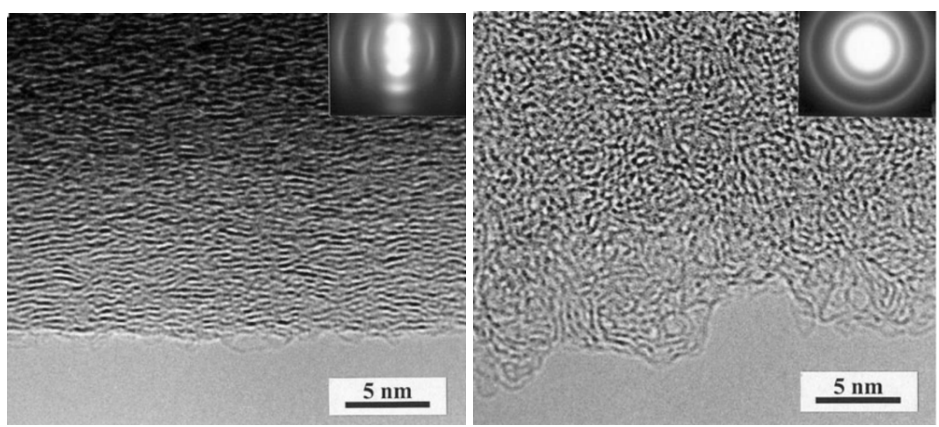

**Figure 6. HRTEM images of graphitising (left) and non-graphitising (right) materials, insets show the diffraction pattern of the area. Reprinted from Philos. Mag. Lett., 80, Harris, P. J. F., Burian, A. & Duber, S. *High-resolution electron microscopy of a microporous carbon*, 381–386, Copyright (2000), with permission from Taylor & Francis Ltd.[72]**

The morphologies and nanostructures within biochars can be challenging to define in absolute terms. As a result, many techniques used to characterise these nanostructures yield only qualitative or relative results. These are, nevertheless, incredibly useful when attempting to understand biochars on the molecular level.

High-Resolution Transmission Electron Microscopy (HRTEM) can be used to image nano-scale regions of a biochar sample and can produce micrographs with up to atomic-level resolutions. From these, a wide range of features can be identified including pentagonal, hexagonal and heptagonal rings, graphitic and non-graphitic regions and a range of different nanostructures.[8,11,47,55,82,84–86] However, due to the extremely high magnifications used, sample inhomogeneity can heavily bias the results of a HRTEM analysis. To reduce this, many areas of the sample should be imaged to obtain a representative overview of the sample and its nanostructures.

It is also possible to use HRTEM to visualise individual components of a biochar by first carrying out solvent extraction and liquid phase exfoliation.[47,64,68] While this provides exciting and specific insights into the molecular structures, the methods through which individual components are removed are biased towards those which can be displaced from the main sample. As with MS, these components may not be representative of the overall material and analysis of these alone may yield a biased view.

Raman spectroscopy is often used in conjunction with TEM[8,11,55,84,85] to detect *Raman active* vibrational modes (those which scatter radiation inelastically) within a sample. Within Raman spectra, the presence of a peak centred around 1590 cm$^{-1}$ can be attributed to *ordered graphitic* nanostructures, such as those of crystalline graphite, and the presence of a broad peak centred around 1350 cm$^{-1}$ can be attributed to *defective* or *disordered* nanostructures, such as those present in non-graphitising materials.[55,84,85,87] These two peaks are referred to as G and D peaks, respectively. Simply the presence of a D peak in the Raman spectrum of a biochar indicates disorder within its nanostructures, yet, the relative extent of disorder in a sample can also be assessed through comparison of the D and G peak areas[11,88–91] or peak intensities.[30,64,84,85,89,92,93] A low D to G ratio indicates a relatively low degree of disorder, whereas a high D to G ratio indicates a relatively high degree of disorder.[55,74,77,89,91] This D to G ratio gives relative information about the nanostructures within a biochar and can be used to compare between samples, but not to obtain absolute measures. However, as both peak areas and peak intensities, along with a range of different deconvolution procedures, are used throughout the literature, many D to G ratios are not directly comparable.[69,84,87,89–91,94,95] Furthermore, excessive grinding of a sample, which may mechanically introduce defects and add to the D peak,[88,95] or unintentional heating of the sample by the incident radiation, which may lead to peak shifting, can also be sources of error and uncertainty when characterising biochars using Raman spectroscopy.[91]

Despite biochar's tendency to resist graphitisation, short-range graphitic structures have been found within them.[11,23,47,55,84,96] These graphitic nanostructures are thought to consist of nanoscale regions of π-π stacked aromatic sheets, often referred to as *crystallites*.[54] When analysed by X-ray diffraction (XRD) or electron diffraction (ED), crystallite-containing biochars produce diffractograms characteristic of crystalline graphite.[9,23,45,46,55,81,96–98] The inserts of the **Figure** 6 are the Selected Area Electron Diffraction (SAED) patterns, carried out alongside TEM. These patterns demonstrate the symmetric diffraction rings of amorphous structure and arcs, confirming layer alignments for the graphitizing structure, with the short sharp peaks corresponding to the {002} *interlayer spacings*. The diffraction pattern, is an overlay of signals, which can be deconvoluted through Fourier Transform and then used to calculate the *interlayer spacing*, *average domain size* and *stack height* of crystallites within the sample (**Figure 7**).

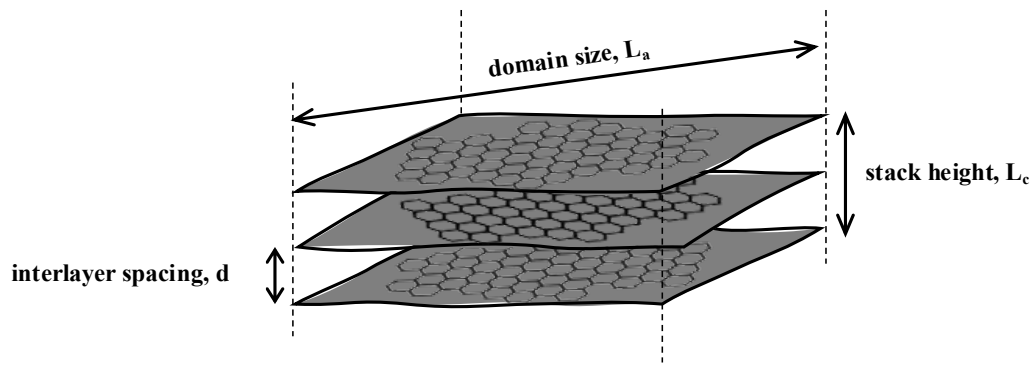

**Figure 7. A schematic guide to the interlayer spacing, domain sizes and stack heights of graphitic crystallites within biochars.**

The Bragg's law defines a geometric relation between atomic planes and the scattering angles at a given wavelength,[99,100] and so, it allows to calculate the interlayer spacings within the crystalline material. Knowing the X-rays wavelength allows for determining the atomic distances, producing sharp peaks for an orderly crystal. Nevertheless, stacking faults and disordered materials will result in a broader, overlapping peaks with lower intensity, creating the uncertainties in the of the measured distances.

On the other hand, the *average domain sizes* and *stack heights* of crystallites are calculated using the Scherrer equation, which is based on the idea that crystallites of < 200 nm cause peak broadening in a diffractogram.[101] However, there is some debate as to whether this holds true in complex carbonaceous materials, as the Scherrer equation does not account for the many other factors that can also contribute to this peak broadening.[102] For example, instrumental effects, curvature caused by strain within the crystallites and defects, such as dislocations and grain boundaries, within the nanostructures of a sample.[102] In biochars, it is near impossible to determine which of these many factors is the main source of peak broadening. Furthermore, the Scherrer equation uses a shape constant, *K*, to calculate the *average domain sizes* and *stack heights* within a biochar sample.[101] This constant must be chosen to represent the shapes of fragments within crystallites, yet, for biochars, this shape is unknown. Many different values of *K* have been suggested but a consensus has yet to be reached.[37,45,103–106] As a result, the calculated *average domain sizes* and *stack heights* will vary depending on the chosen $K$ value. Other causes of uncertainty in these calculated values may arise due to peak asymmetry, noise/background scattering and/or issues when fitting curves to peaks.[37]

These techniques give complementary insights into the aromatic molecular structures within biochars and can be combined to gain an overview of the nanostructures present within a sample.

## 2.4. Functional groups

Biochars contain significantly fewer functional groups than their biomass precursors. These are related, through chemical transformations, to those found within biomass. Carbon and oxygen-based functionalities dominate, including aryl, hydroxyl, carbonyl and many more. Examples of these are shown in **Figure 8**.[1,9,96,107–115,12,116–118,16,22,25,36,39,49,92] Depending upon the composition of the starting biomass, a variety of nitrogen-, phosphorous- and sulphur- based functionalities may also be present, nevertheless those are found in significantly lower quantities than oxygen-based groups.[116,119]

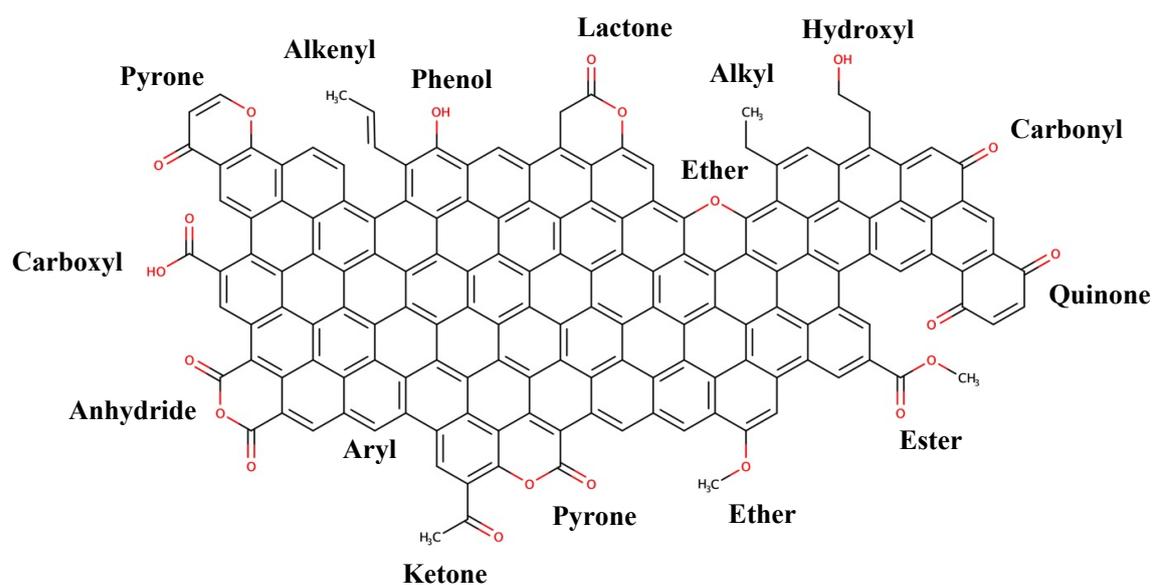

Figure 8. Examples of types of carbon- and oxygen-based functionalities present within biochars.

Depending on its origin, the biomass may contains a wide variety of functional groups, nevertheless many become unstable during pyrolysis. As a result, the range of functional groups present within biochars are much more limited. The approximate stabilities of a variety of common oxygen-based functional groups are shown in **Figure 9**. From this, we can see that each functional group decomposes across a wide temperature range, as opposed to at a single value. This is due to the influence of the chemical environment surrounding a group. This effect is particularly of note in the production of biochars, as a range of factors will affect how functional groups develop during pyrolysis. For example, as previously mentioned, inorganic compounds can have both catalytic and stabilising effects, and so the presence of these compounds may significantly effect decomposition and/or transformations of functional groups when producing a biochar.[12] Feedstock inhomogeneity may exacerbate this effect and result in significant variation in the functional groups of the produced biochar.[120] Nevertheless, from **Figure 9**, we can see that certain functional groups (e.g. carbonyls and pyrones) are more stable than others (e.g. carboxyl). At the same time, carboxylic group will decompose 100 degrees earlier when in a highly acidic form, than its weakly acidic counterpart where H-bonding with neighbouring atoms will be stabilising it.[121]

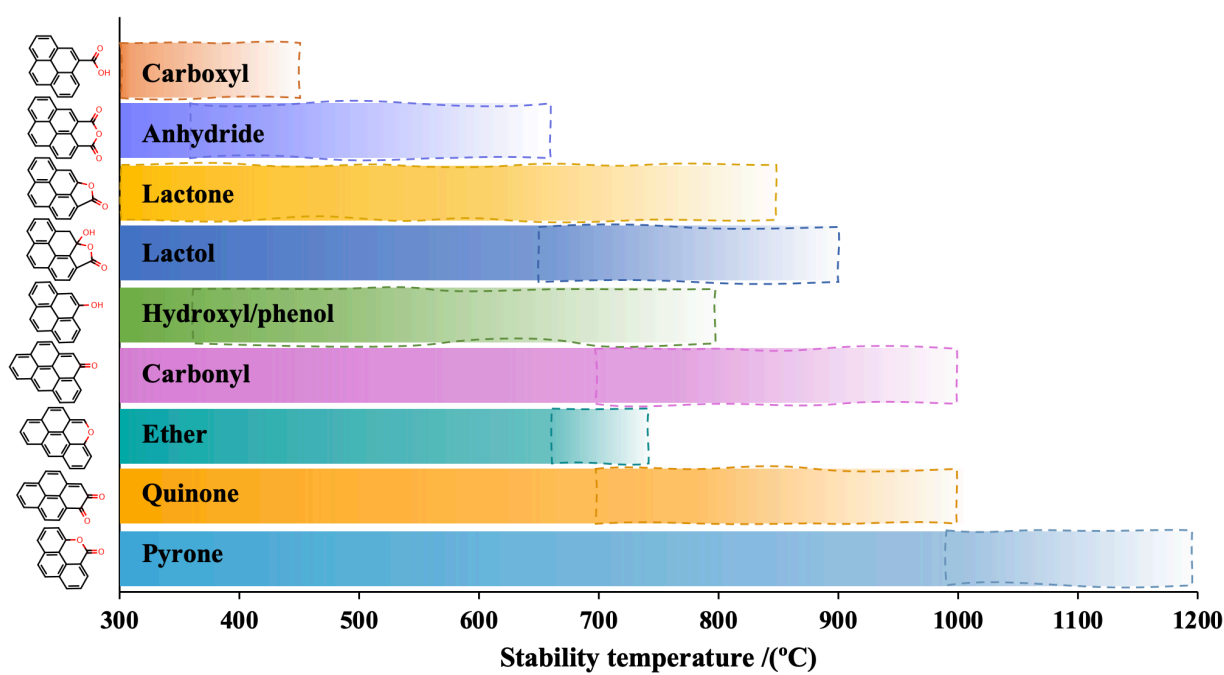

Figure 9. A schematic guide to the stability of oxygen-containing functional groups. Coloured bars indicate temperatures at which each functional group is stable and dashed areas indicate approximate temperature ranges at which each functional group becomes unstable. Refs 121–129.

Despite only being present in relatively low numbers, the functional groups within biochars are central to how biochar interact with their surrounding environment. A holistic understanding of these functionalities, particularly those exposed on surfaces, is, therefore, crucial when building a molecular-level understanding of these materials. A combination of analytical techniques can be used to achieve this.

Fourier Transform Infrared (FTIR) spectroscopy is most commonly used to characterise the functional groups within biochars by probing their vibrational frequencies.[1,9,113,114,12,22,25,92,96,110–112] This is typically done by detecting IR frequencies reflected by the surface of a sample.[25,36,37,130–132] However, low resolutions and difficulties in assigning peaks, particularly when broad or overlapping peaks are present, can be problematic when attempting to characterise the functional groups using FTIR. These spectral features make it difficult to confidently distinguish between functionalities and so only a loose identification of functional group is often possible.

X-ray Photoelectron Spectroscopy (XPS) also offers insights into the functional groups present within biochars by probing atomic environments upon the surface of a sample, typically up to ~10 nm depth.[16,36,115–117] The relative quantities of $sp^2$ and $sp^3$ carbons, aromatic carbons and single and double-bonded oxygens within a sample can be identified in this way. As a highly sensitive surface technique, XPS is strongly influenced by contamination, oxidation or the presence of moisture upon the surface being examined, and so samples should be prepared in situ to yield accurate results.[36]

The acidic functional groups within biochars can be distinguished through their dissociation constants ($pK_a$'s). Strongly acidic groups, such as carboxylic acids and anhydrides, can be neutralised with weak bases, whilst mildly acidic groups, such as lactones, and weakly acidic groups, such as hydroxyls and phenols, require progressively stronger bases.[36,107–109]

Boehm titrations utilise these differences in acidity to determine the amounts of strongly acidic, mildly acidic and weakly acidic functional groups present upon the surface of a sample which, in biochars, are assumed to be carboxyl, lactonic and phenolic functional groups respectively.[16,36,39,49,118,133] The method is, in principle, straightforward – the carbon material is first treated with a reaction base, which is neutralized by the most acidic functional groups, second the remaining base is quantified by the acid-base titration. The standard bases from $pK_a$=6.4 ($NaHCO_3$) to $pK_a$=20.6 (NaOEt) are used, and are assumed to neutralize all oxygens that are more acidic.[108] These titrations were originally developed to analyse carbonaceous materials with negligible ash content,[107–109] an so will be affected by the presence of ash within a sample due to the pH-altering effects of soluble inorganic species.[36] Ash and any other buffering materials should, therefore, be removed prior to analysis, yet, this is infrequently carried out and may lead to incorrect results.[16,39,49,118,133]

When characterising biochars, FTIR, XPS and Boehm titrations will only detect functional groups which are either exposed (or very close) to the samples surface. The extent of sample grinding prior to analysis likely influences results. Although, finely ground samples offer the most complete overview of the functionalities present within the overall biochar, the grinding process may mechanically alter the molecular structures within a sample and lead to skewed results.[88,95,134–137] On the other hand, analysis of unground samples may yield unrepresentative results and can be heavily influenced by sample inhomogeneity. A combination of the above techniques should be used to gain complementary insights into the functionalities present within a biochar sample.

## 2.5. Point of net zero charge and pH

Adding a biochar to a solution will cause a change in pH due to the presence of minerals, soluble inorganic species, and the protonation/deprotonation of acidic/basic functional groups at the surface. In solution, these groups will protonate and deprotonate in accordance with their $pK_a$'s, allowing the biochar surface to become positively or negatively charged.

Two pH-related properties are often measured in biochar research: the '*pH*' of the biochar, which gives an indication of the overall acidity or basicity of a sample and is measured as the pH at which a solution containing biochar equilibrates; and the *point of net zero charge* ($pH_{pzc}$), which is a measure of the pH at which the net surface charge of a biochar is zero, and is calculated as the pH at which the addition of biochar to solution results in no change in pH.[12,26,27,36,42,49,110] However, both measures are dictated by ash and soluble inorganic species and give relatively limited insight into the functional groups present within the organic fraction of a biochar.[12,16,26,49,138] Furthermore, both 'pH' and $pH_{pzc}$ are heavily dependent on experimental set-up and the lack of a standardised protocols renders many literature values incomparable.

## 2.6. Density and porosity

Biochars are mesoporous materials, with pore size distributions spanning multiple length scales. These pores are divided into three categories based on their diameters: *macropores* with diameters between 0.05 and 1000 μm, *mesopores* with diameters between 2 and 50 nm and *micropores* with diameters less than 2 nm.[1] Generally, the *macropores* within a biochar sample are determined by the structures of the feedstock.[16,36,97,139] On the other hand, meso- and micro-porosity develop throughout the pyrolysis process as volatile gases escape the solid matrix.

The density of porous materials can be defined in several ways depending upon the extent to which pores are included in the volume measurement. Three density measures are relevant to the study of biochars: *bulk density*, *envelope density* and *true density*. These densities and their corresponding volumes are depicted in **Figure 10**.

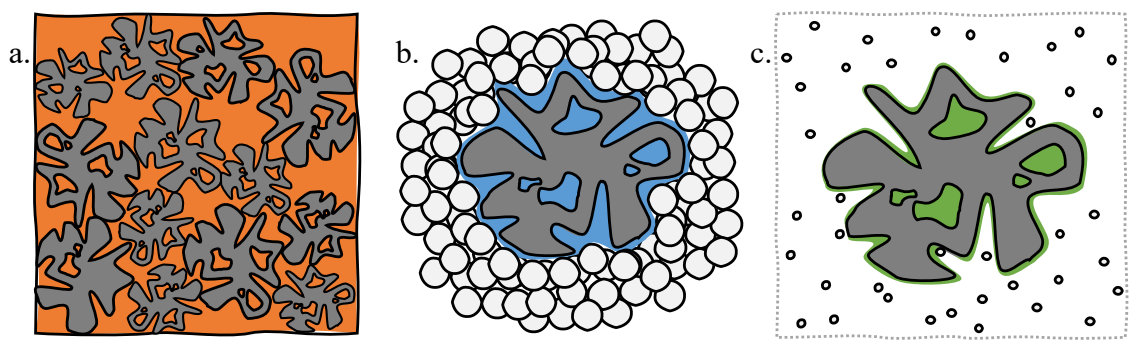

**Figure 10.** Schematic guide to the densities of biochars (in grey): (a) shows the volume measured (in orange) when calculating tap density, (b) shows the volume measured (in blue) when calculating envelope density, and (c) shows the volume measured (in green) when calculating true density.

Bulk density, also known as *packing density* or *tap density*, is the simplest measure of density. It is found by adding a known sample mass to a container and measuring its volume after vibrating or tapping to obtain optimal packing of particles. Bulk densities include the volumes of the particles themselves, the pores within them (*intra-particle pores*) and of the voids between them (*extra-particle voids*) (**Figure 10a**). Therefore, they largely depend on the sizes and shapes of particles within a sample – with those that can pack together more effectively returning higher bulk densities.

Envelope density, also known as *apparent density* or *particle density*, is a more refined measure of density. It represents the average density of individual particles within a sample, including intra-particle pores whilst excluding extra-particle voids. The envelope density of a sample can be calculated by measuring the volume displaced when a known mass of the sample is completely immersed and consolidated in a container of smaller non-wetting particles.[36,44] In this way, the volumes of only the biochar particles and their intra-particle pores are included in the calculation of envelope density (**Figure 10b**). When carrying out this measurement, it is important that the non-wetting particles completely surround and pack tightly around the sample. The sizes of non-wetting particles, the forces used to consolidate them and the ratio of sample to non-wetting particles can therefore significantly impact results.[44] The non-wetting particles should be small enough, the consolidation force should be high enough, and the ratio of sample to non-wetting particles should be low enough to ensure all extra-particle voids are occupied. However, consolidation forces which are too large may cause the fracturing of particles and should, therefore not be excessively high.[44]

True density, also known as *skeletal density* or *helium density*, is the finest measure of density. It represents the average density of individual particles within a dried sample, excluding intra-particle pores and extra-particle voids. True densities are measured using helium pycnometry. This technique uses the isothermal expansion of helium gas to calculate the volume occupied by a sample. It is assumed that helium gas can penetrate all open/accessible intra-particle pores and, therefore, that only the volumes of the particles themselves and any closed/inaccessible intra-particle pores are included in the calculation of true density (**Figure 10c**). This is an important assumption when calculating true densities, as changes to particle sizes (e.g. through grinding) can expose additional intra-particle pores and thereby increase the measured true density of the sample.[44] Fine-grinding prior to analysis can reduce this variability, by exposing the maximum number of intra-particle pores, thus allowing comparison between samples.[44] Degassing to remove adsorbed moisture, volatile compounds and/or tarry substances, which can also block pores and lead to error, is also recommended.[36] Furthermore, as the closest approach of a helium atom to the surface of a sample is its collision radius (~1.3 Å),[140] a volume near the surface will be excluded from analysis.[140] In materials with very high specific surface areas (SSAs), such as activated carbons which have SSAs on the order of thousands of $m^2g^{-1}$, this excluded-volume effect can cause significant errors in results.[140–142] However, it has been shown to be negligible in biochar-like materials which have SSAs only on the order of hundreds of $m^2g^{-1}$.[140,142] A second important assumption made when calculating true densities by helium pycnometry is that helium does not adsorb to the samples surface. This, however, is not the case for all carbonaceous solids.[141,143,144] Adsorption effects become especially significant in high SSA materials[140,141,144] and so, as with excluded-volume effects, adsorption effects have been shown to be negligible in biochars.[144] Further sources of error or uncertainty in these measured true densities can arise from insufficient equilibration or measurement cycles and, sample inhomogeneity.[145] Generally, using 20+ equilibration cycles and averaging across 10+ measurement cycles is therefore recommended.[15,34,36,37,145]

True densities are of most relevance when attempting to understand biochars on the molecular level as they include only minimal porosity and therefore give the best indication of the density and packing of nanostructures within a sample. Comparison of the true densities of biochars to that of graphite can yield insights into the degree of aromatic condensation and graphitisation of biochar: in general, biochars have densities below that of graphite, indicating they have lower degrees of aromatic condensation and less ordered nanostructures. The measured true density of a biochar, however, is an average of all different components within it, including both organic and inorganic fractions. This should be considered when using true densities of biochars to obtain information about their nanostructures. In general, the inorganic fraction of a biochar will have a higher density than the organic fraction[34] and so, in high ash biochars, the measured true densities will be slightly inflated.

When combined, envelope densities and true densities can also be used to gain insights into the porosity of a biochar as a percentage volume.[44,146] This porosity is calculated as the percentage difference between the densities measured by each technique and therefore represents the percentage volume of intra-particle pores within a biochar sample. However, this calculation is only valid when both measurements are carried out on samples with approximately equivalent particle sizes.[36]

A second technique used to characterise the densities and porosities of biochars is mercury porosimetry. Mercury is a non-wetting fluid and so will only enter the pores of a material under pressure. During a mercury porosimetry measurement, a sample is enveloped by mercury and the pressure of the system is gradually increased, forcing the mercury to enter progressively smaller pores. The volume changes are measured throughout this process, allowing the porosity and pore size distributions (PSD) within a sample, along with its envelope and true density, to be calculated.[44] Mercury porosimetry can be problematic for biochars, as the high pressures may cause damage to the sample, leading to permanent changes in its pore structures and erroneous in results. Furthermore, although pores are assumed to be cylindrical when calculating PSDs, a range of different pore shapes/structures are likely to be present within biochars and the diameter of pore openings may not be representative of internal pore diameters. As a result, the calculated PSDs may differ from those within the sample. In addition, the contact angle between the mercury and the sample, which is used to calculate porosity and PSDs, is likely to vary greatly due to surface roughness and surface-exposed functional groups yet, this is infrequently considered.[11,44,147]

## 2.7. Surface area and porosity

Biochars have relatively high specific surface areas (SSAs) due to their porous nature. Gas adsorption is traditionally used to characterise these SSAs, with $N_2$ and $CO_2$ being the most common choices of the adsorbate.[22,30,149–153,34,36,44,49,82,96,139,148] Gas adsorption techniques generally rely on applying adsorption models to calculate the SSAs of a sample of known mass from the relationship between adsorbed amount and adsorbate pressure. Both the choice of adsorption model and adsorbate will affect these calculated properties.[37,149,151] Experimental set up can also significantly affect the adsorption isotherms of biochars. Degassing temperature, for example, has been shown to have strong effects due to the volatisation of organic and tarry compounds at increased temperatures.[151]

The Brunauer-Emmett-Teller (BET) model is most commonly used to calculate the SSAs of biochars.[22,34,49,96,149,151,153] It assumes all adsorption sites are equivalent, and so, it is unsuitable for characterisation of heterogenous or microporous materials.[148,154] Biochars, however, are both heterogenous and microporous, which makes the BET model prone to incorrect result. Furthermore, the standard relative pressure range of 0.05 to 0.35 is often too large when applied to biochars and should be adjusted to ensure validity of the BET model.[148,149,155,156] Although methods have been developed to aid in this adjustment,[148,155] the correct relative pressure range can be difficult to identify,[148,149] leading to further error and uncertainty.

The adsorption of $N_2$ at 77 K is typically used to determine the SSAs of biochars. When using the BET model, the SSA of the sample is found by multiplying the molecular cross-sectional area of $N_2$ by the monolayer adsorbed amount and normalising by the mass of the sample being analysed. As standard, the molecular cross-sectional area of $N_2$ is calculated assuming molecules are adsorbed in a hexagonal close-packed monolayer.[148,154] However, $N_2$ has a quadrupole moment which can interact with polar surfaces, such as those of biochars, and lead to deviations from this standard value. As a result, BET SSAs of biochars calculated using the standard $N_2$ cross-sectional area may be significantly incorrect.[148,154] Hysteresis, long equilibration times and micropore filling/condensation can lead to further errors. These issues may be overcome through the use of alternative adsorption models and/or alternative adsorbate gases.[148,154]

Gas adsorption can also be used to assess pore volumes and PSDs within biochars.[30,44,82,139,149–152] As with SSAs, this can be achieved through the application of adsorption models, such as the Barrett-Joyner-Halenda (BJH) model,[30,98,132,139,148,157] or through computational techniques, such as Density Functional Theory (DFT) and non-local DFT (NLDFT),[148–150,158] which use fluid dynamics to characterise pores.

$N_2$ at 77 K is kinetically unable to access micropores < 0.5 nm, instead, $CO_2$ at 273 K is often used when characterising the pore volumes and PSDs of biochars.[148,149,156] $CO_2$ isotherms can be analysed in much the same way as $N_2$ isotherms, however, as $CO_2$ is only able to probe pores up to approximately 1.5 nm, only microporosities can be characterised using $CO_2$ adsorption isotherms.[148] Unfortunately, $CO_2$ possesses an even greater quadrupole moment than $N_2$, and so may still yield unreliable results.[148,151,154,159] Gases such as Ar, $O_2$ and $H_2$, which have much smaller (or zero) quadrupole moments, have been proposed as alternatives and are likely, in future, to replace $N_2$ and $CO_2$ in the characterisation of biochars and biochar-like materials.[148,154,159160,161]

## 3. Highest treatment temperature (HTT) effect on biochar structure and properties

Although many conditions can be varied during biochar production, the highest treatment temperature (HTT) reached during pyrolysis is widely accepted as the most influential of these.[27] As a result, thermo-sequences, in which biochars are produced across a series of HTTs, are often used when studying their properties. The temperature ranges studied generally range from as low as 200 – 300ºC and reach highs of 1000 – 1200ºC. Thus, the changes in biochar properties with increasing carbonisation can be seen.

We collected data from the UKBRC Charchive (http://www.charchive.org/), UC Davis Biochar Database (http://biochar.ucdavis.edu/) and a number of published studies,[2,15,162,163,38,44,48,56,58–60,146] in order to observe these effects, focusing solely upon biochars produced from woody feedstocks. The trends observed through this data are also true of other feedstock types.[2,12,16,25–27,58,164,165] The collected data can be downloaded from github.com/Erastova-group/Biochar_MolecularModels.

The influence of HTT on proximate composition is shown in **Figure 11** and ultimate composition is shown in **Figures 12** and **13**. From these, we can clearly see that with increasing HTT, the fixed carbon (FC) content of biochars increases whilst the volatile matter (VM) content and, H/C and O/C molar ratios decrease. These changes correspond to the increase in carbonisation of the organic matter within biochars with increasing HTT *via* the formation of stable carbon-rich molecular structures and the loss of oxygen- and hydrogen-rich volatile organic compounds. The effects of HTT on moisture and ash content (**Figure 11c-d**), however, are less obvious as these components comprise much smaller fractions of the total biochar and are, therefore, more strongly influenced by the abovementioned characterisation issues. In general, moisture slightly decreases with HTT, whereas ash becomes relatively enriched with increasing HTT due to the loss of organic matter.

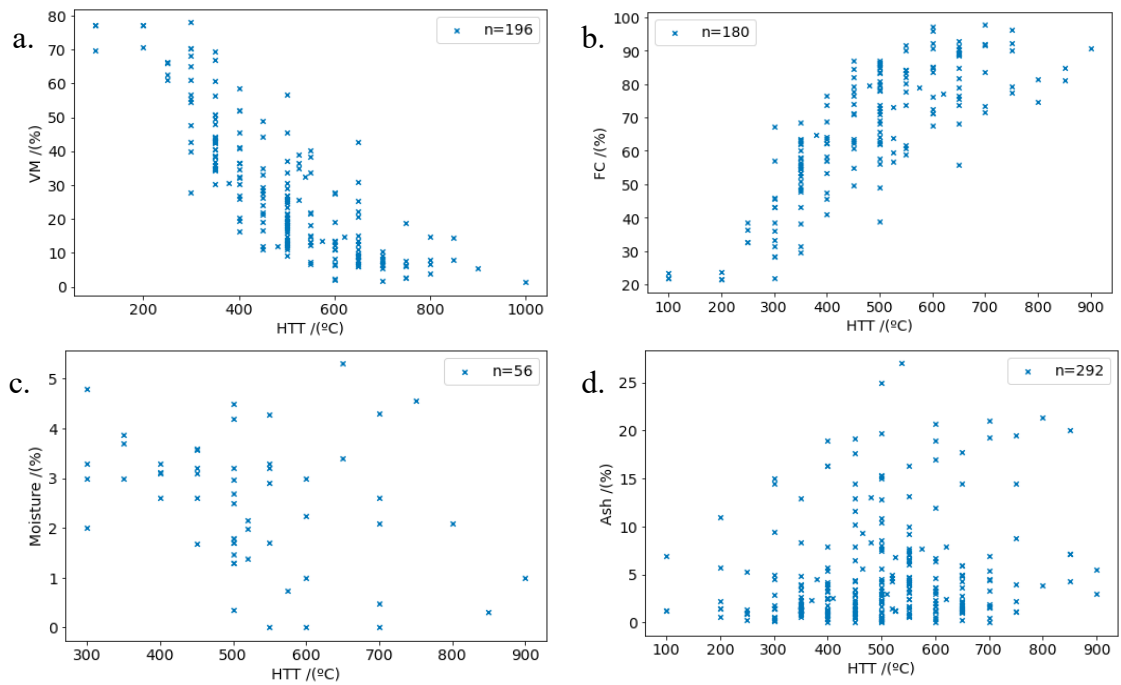

**Figure 11. Influence of highest treatment temperature (HTT) on proximate composition of woody biochar, showing: (a) volatile matter (VM), (b) fixed carbon (FC), (c) moisture, and (d) ash as weight percent. The number of data points, n, for each plot is given in the legend.**

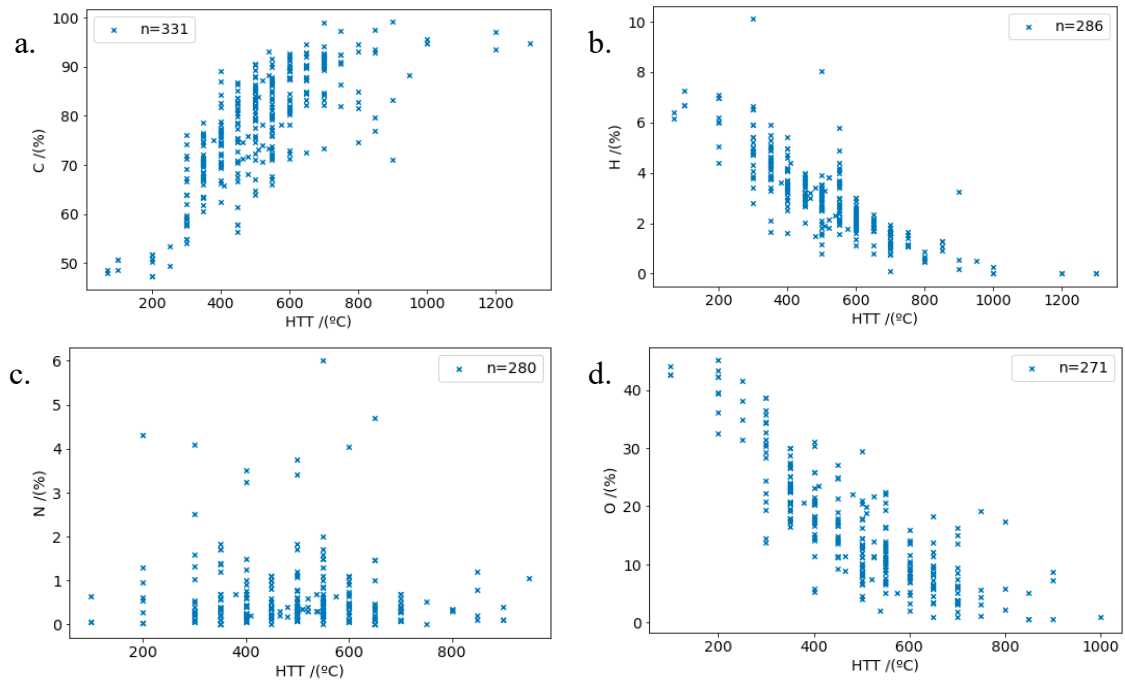

**Figure 12. Influence of highest treatment temperature (HTT) on ultimate composition of woody biochar, presenting: (a) C, (b) H, (c) N, and (d) O contents as weight percent. The number of data points, n, for each plot is given in the legend.**

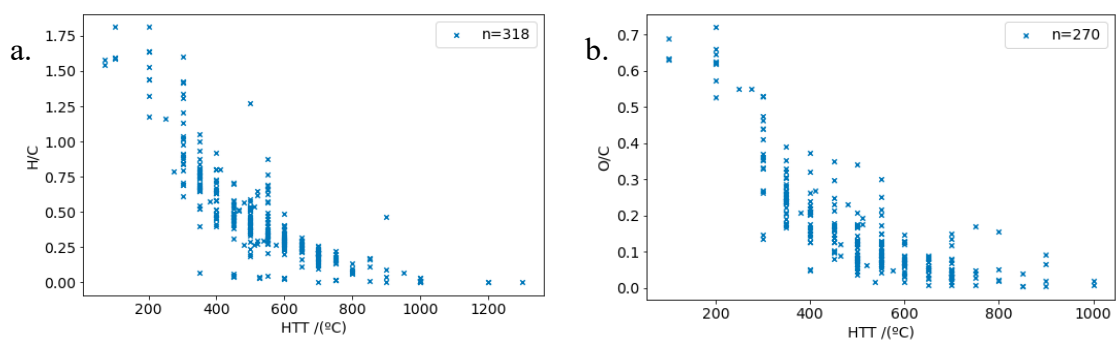

**Figure 13. Influence of highest treatment temperature (HTT) on (a) H/C and (b) O/C molar ratios of woody biochars. The number of data points, n, for each plot is given in the legend.**

**Figure 14** shows the influence of HTT on *aromaticity index* and *aromatic domain size*. From this, we can see that both *aromaticity index* and *aromatic domain size* increase with increasing HTT. Whilst *aromaticity index* increases fairly rapidly between 200-500ºC, *aromatic domain size* increases more gradually. This corresponds to the increase in carbonisation of the organic matter within biochars with increasing HTT, first through the formation of smaller clusters of aromatic rings and then through the gradual condensation of these structures into a highly conjugated aromatic carbon network.

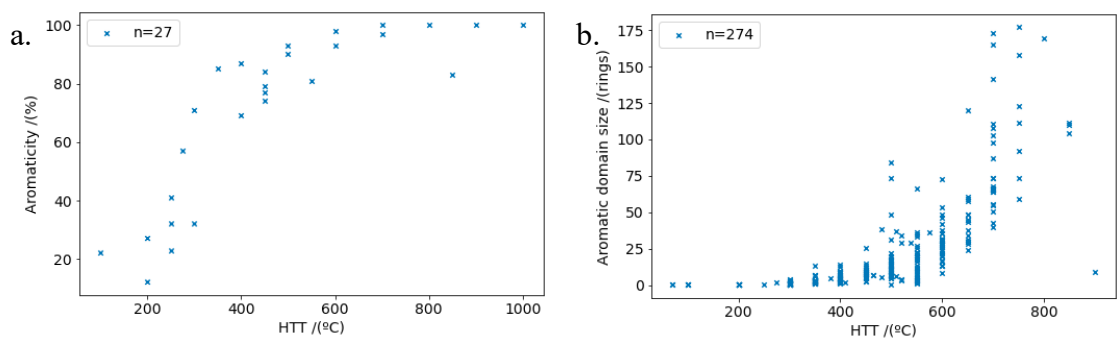

**Figure 14. Influence of highest treatment temperature (HTT) on (a) aromaticity and (b) aromatic domain size of woody biochars. The number of data points, n, for each plot is given in the legend.**

The influence of HTT on pH is shown in **Figure 15**. From this, we can see that pH generally increases with increasing HTT. Whilst this change is primarily due to the increase in alkaline ash content (**Figure 11d**), it is also partly related to the loss of acidic functional groups (**Figures 9** and **12b**) from the organic matter of these biochars with increasing HTT.

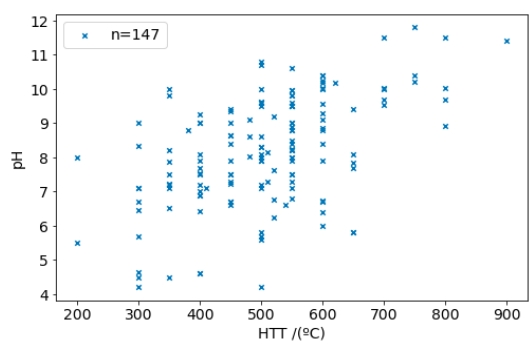

**Figure 15.** Influence of highest treatment temperature (HTT) on 'pH' of woody biochars. The number of data points, n, for each plot is given in the legend.

The influence of HTT on *true density* and specific surface area (SSA) is shown in **Figure 16**. From this, we can see that *true density* increases with increasing HTT and plateaus at a density of ~2000 kg m$^{-3}$, just below that of graphite (density ~2250 kg m$^{-3}$).[14,15,48] Again, this change primarily corresponds to the increase in carbonisation of the organic matter of these biochars with increasing HTT. Surface area also appears to increase with increasing HTT, however the larger spread of data caused by the issues outlined earlier and the lack of data points at higher HTTs makes this trend more challenging to discern. The increase in SSA with increasing HTT is due to the development of porosity, particularly *micro-* and *meso-* porosity, at increased HTTs. This porosity is formed as volatile compounds exit the biochar, leaving behind voids and channels and exposing additional surface.[164]

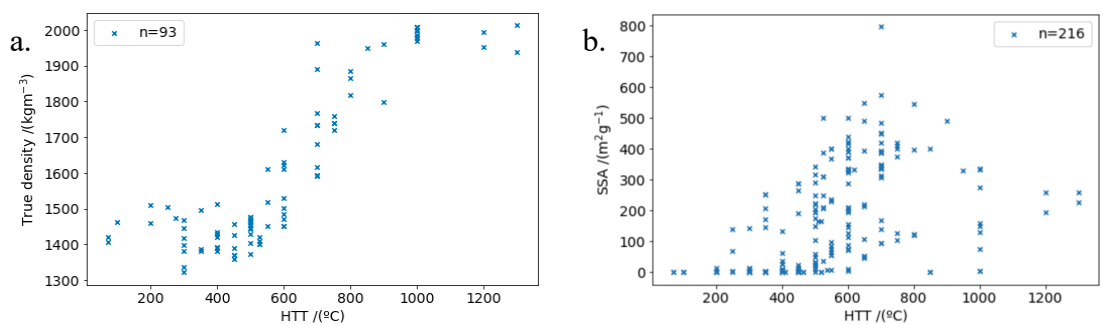

**Figure 16.** Influence of highest treatment temperature (HTT) on (a) true density and (b) specific surface area (SSA) of woody biochars. The number of data points, n, for each plot is given in the legend.

The influence of HTT on the functional groups within biochars is difficult to quantify numerically. However, these effects can still be observed and understood through a review of the literature. In woody biochars, functional groups will be predominantly oxygen-based due to the woody feedstocks' low nitrogen, phosphorus and sulphur contents.[12,26,96,116,119,164] Biochars produced at lower HTTs have lower *aromaticity indices* and, therefore greater numbers and a wider variety of functional groups. With increasing HTT, both hydrogen and oxygen are lost

(**Figure 12** and **13**) and, as a result, the total numbers of functional groups present will decrease. Furthermore, as many functional groups become unstable at high temperatures (**Figure 9**), the range of functional groups present will be narrowed.[25] Biochars produced at high HTTs will be predominantly aromatic, with limited numbers of more thermally-stable functional groups, many of which will be conjugated into the aromatic molecular structures of these materials.[25]

# 4. Conclusion

The aim of our work is to bring much-needed molecular-level understanding of biochar materials. Here we present the first part of the work, where we have critically assessed the analytical techniques used to characterise biochars and, in doing so, assembled the information necessary to develop a molecular-level understanding of these materials. We have gathered and presented a large collection of characterisation data, obtained from public domains; allowing us to gain insights into the changes in the chemical, physical and molecular properties of woody biochars as a function of biochar's processing, quantified as the highest treatment temperature. This information, supported by an understanding of limitations of each analytical technique, has allowed us to gain understanding of the molecular structures comprising woody biochars and thus, paved the way for the next part of this work – the development of realistic biochar molecular models for simulations. Further analysis of the collected data and the development of a set of realistic biochar models for molecular simulations can be found in the Wood *et al.*, "Biochars at the molecular level. Part 2 – Development of realistic molecular models of biochars."[3]

# Acknowledgements

Rosie Wood would like to thank E4 DTP for the funding of the PhD project "Molecular Modelling for Design of Biochar Materials". Valentina Erastova would like to thank Chancellor's Fellowship by the University of Edinburgh.

# TOC

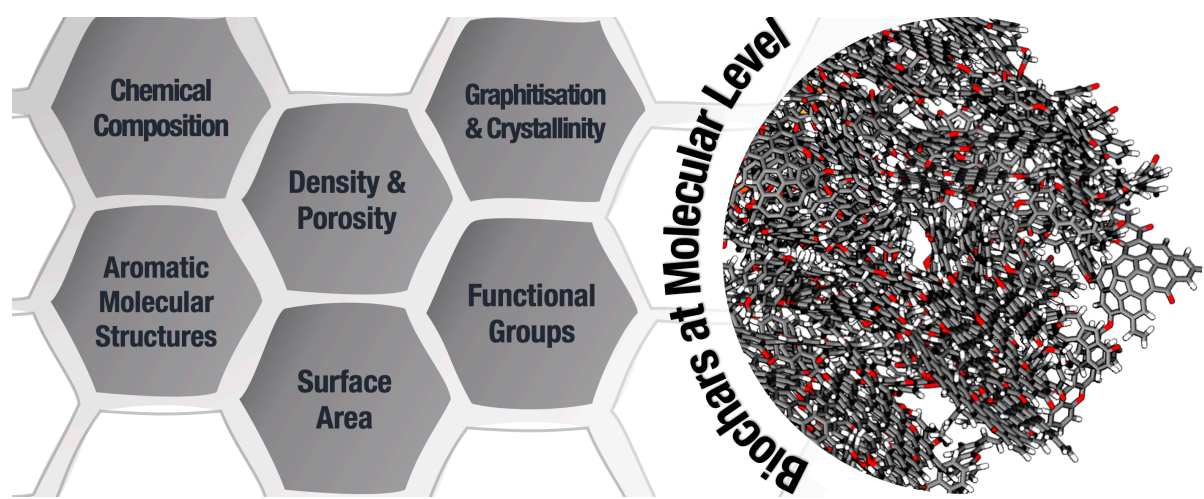

# References

1. Lehmann, J., Joseph, S. & Joseph, S. *Biochar for Environmental Management. Biochar for Environmental Management* (Taylor & Francis, 2015). doi:10.4324/9780203762264.

2. Weber, K. & Quicker, P. Properties of biochar. *Fuel* **217**, 240–261 (2018).

3. Wood, R., Masek, O. & Erastova, V. Biochars at the molecular level. Part 2 -- Development of realistic molecular models of biochars. (2023).

4. Burnham, A. K. *Global Chemical Kinetics of Fossil Fuels: How to Model Maturation and Pyrolysis. Global Chemical Kinetics of Fossil Fuels: How to Model Maturation and Pyrolysis* (Springer International Publishing, 2017). doi:10.1007/978-3-319-49634-4.

5. Speight, J. G. *Handbook of Coal Analysis. Handbook of Coal Analysis* (John Wiley & Sons, Inc, 2015). doi:10.1002/9781119037699.

6. KREVELEN, V. & W., D. Graphical-statistical method for the study of structure and reaction processes of coal. *Fuel* **29**, 269–284 (1950).

7. Krevelen, D. W. Van. *Coal, Third Edition*. (Elsevier, 1993).

8. Surup, G. R. *et al.* The effect of feedstock origin and temperature on the structure and reactivity of char from pyrolysis at 1300–2800 °C. *Fuel* **235**, 306–316 (2019).

9. Smidt, E. *et al.* Infrared spectroscopy refines chronological assessment, depositional environment and pyrolysis conditions of archeological charcoals. *Sci. Rep.* **10**, 1–11 (2020).

10. Bakshi, S., Banik, C. & Laird, D. A. Estimating the organic oxygen content of biochar. *Sci. Reports 2020 101* **10**, 1–12 (2020).

11. Surup, G. R., Nielsen, H. K., Heidelmann, M. & Trubetskaya, A. Characterization and reactivity of charcoal from high temperature pyrolysis (800–1600 °C). *Fuel* **235**, 1544–1554 (2019).

12. Domingues, R. R. *et al.* Properties of biochar derived from wood and high-nutrient biomasses with the aim of agronomic and environmental benefits. *PLoS One* **12**, e0176884 (2017).

13. Pereira, B. L. C. *et al. Wood chemistry & charcoal. BioResources* vol. 8 (2013).

14. Brewer, C. E. & Brown, R. C. Biochar. in *Comprehensive Renewable Energy* vol. 5 357–

384 (Elsevier, 2012).

15. Tintner, J. *et al.* Impact of Pyrolysis Temperature on Charcoal Characteristics. *Ind. Eng. Chem. Res.* **57**, 15613–15619 (2018).

16. Suliman, W. *et al.* Influence of feedstock source and pyrolysis temperature on biochar bulk and surface properties. *Biomass and Bioenergy* **84**, 37–48 (2016).

17. Kelemen, S. R., Afeworki, M., Gorbaty, M. L. & Cohen, A. D. Characterization of organically bound oxygen forms in lignites, peats, and pyrolyzed peats by X-ray photoelectron spectroscopy (XPS) and solid-state 13C NMR methods. *Energy and Fuels* **16**, 1450–1462 (2002).

18. Zhao, C., Jiang, E. & Chen, A. Volatile production from pyrolysis of cellulose, hemicellulose and lignin. *J. Energy Inst.* **90**, 902–913 (2017).

19. Yang, H., Yan, R., Chen, H., Lee, D. H. & Zheng, C. Characteristics of hemicellulose, cellulose and lignin pyrolysis. *Fuel* **86**, 1781–1788 (2007).

20. Basu, P. Pyrolysis. in *Biomass Gasification, Pyrolysis and Torrefaction* 155–187 (Elsevier, 2018). doi:10.1016/B978-0-12-812992-0.00005-4.

21. Sadakata, M., Takahashi, K., Saito, M. & Sakai, T. Production of fuel gas and char from wood, lignin and holocellulose by carbonization. *Fuel* **66**, 1667–1671 (1987).

22. Sharma, R. K. *et al.* Characterization of chars from pyrolysis of lignin. *Fuel* **83**, 1469–1482 (2004).

23. Bourke, J. *et al.* Do All Carbonized Charcoals Have the Same Chemical Structure? 2. A Model of the Chemical Structure of Carbonized Charcoal. *Ind. Eng. Chem. Res.* **46**, 5954–5967 (2007).

24. Bajpai, P. Physical and chemical characteristics of lignocellulosic biomass. *Lignocellul. Biomass Biotechnol.* 11–24 (2022) doi:10.1016/B978-0-12-821889-1.00001-1.

25. Janu, R. *et al.* Biochar surface functional groups as affected by biomass feedstock, biochar composition and pyrolysis temperature. *Carbon Resour. Convers.* **4**, 36–46 (2021).

26. Enders, A., Hanley, K., Whitman, T., Joseph, S. & Lehmann, J. Characterization of biochars to evaluate recalcitrance and agronomic performance. *Bioresour. Technol.* **114**, 644–653 (2012).

27. Zhao, L., Cao, X., Mašek, O. & Zimmerman, A. Heterogeneity of biochar properties as a function of feedstock sources and production temperatures. *J. Hazard. Mater.* **256–257**, 1–9 (2013).

28. Mašek, O. *et al.* Consistency of biochar properties over time and production scales: A characterisation of standard materials. *J. Anal. Appl. Pyrolysis* **132**, 200–210 (2018).

29. Mészáros, E. *et al.* Do All Carbonized Charcoals Have the Same Chemical Structure? 1. Implications of Thermogravimetry−Mass Spectrometry Measurements. *Ind. Eng. Chem. Res.* **46**, 5943–5953 (2007).

30. Phounglamcheik, A. *et al.* Effects of Pyrolysis Conditions and Feedstocks on the Properties and Gasification Reactivity of Charcoal from Woodchips. *Energy and Fuels* **34**, 8353–8365 (2020).

31. Collard, F. X. & Blin, J. A review on pyrolysis of biomass constituents: Mechanisms and composition of the products obtained from the conversion of cellulose, hemicelluloses and lignin. *Renewable and Sustainable Energy Reviews* vol. 38 594–608 (2014).

32. Ahmad, M. *et al.* Biochar as a sorbent for contaminant management in soil and water: A review. *Chemosphere* **99**, 19–33 (2014).

33. ASTM D 1762-84. Standard Test Method for Chemical Analysis of Wood Charcoal. *ASTM Int.* **84**, 1–2 (2011).

34. Brewer, C. E., Schmidt-Rohr, K., Satrio, J. A. & Brown, R. C. Characterization of biochar from fast pyrolysis and gasification systems. *Environ. Prog. Sustain. Energy* **28**, 386–396 (2009).

35. Illingworth, J., Williams, P. T. & Rand, B. Characterisation of biochar porosity from pyrolysis of biomass flax fibre. *J. Energy Inst.* **86**, 63–70 (2013).

36. Singh, B., Camps-Arbestain, M., Lehmann, J. & CSIRO (Australia). *Biochar: A Guide to Analytical Methods*. *Csiro publishing* (2017).

37. Nzihou, A. *Handbook on characterization of biomass, biowaste and related by-products*. *Handbook on Characterization of Biomass, Biowaste and Related By-products* (2020). doi:10.1007/978-3-030-35020-8.

38. Solar, J. *et al.* Conversion of Injected Forestry Waste Biomass Charcoal in a Blast Furnace: Influence of Pyrolysis Temperature. *Energy and Fuels* **35**, 529–538 (2021).

39. Pereira, R. C., Arbestain, M. C., Sueiro, M. V. & MacIá-Agulló, J. A. Assessment of the surface chemistry of wood-derived biochars using wet chemistry, Fourier transform infrared spectroscopy and X-ray photoelectron spectroscopy. *Soil Res.* **53**, 753–762 (2015).

40. Calvelo Pereira, R. *et al.* Contribution to characterisation of biochar to estimate the labile fraction of carbon. *Org. Geochem.* **42**, 1331–1342 (2011).

41. Vassilev, S. V., Baxter, D., Andersen, L. K. & Vassileva, C. G. An overview of the composition and application of biomass ash. Part 1. Phase-mineral and chemical composition and classification. *Fuel* vol. 105 40–76 (2013).

42. Banik, C., Lawrinenko, M., Bakshi, S. & Laird, D. A. Impact of Pyrolysis Temperature and Feedstock on Surface Charge and Functional Group Chemistry of Biochars. *J. Environ. Qual.* **47**, 452–461 (2018).

43. Lü, J., Li, J., Li, Y., Chen, B. & Bao, Z. Use of rice straw biochar simultaneously as the sustained release carrier of herbicides and soil amendment for their reduced leaching. *J. Agric. Food Chem.* **60**, 6463–6470 (2012).

44. Brewer, C. E. *et al.* New approaches to measuring biochar density and porosity. *Biomass and Bioenergy* **66**, 176–185 (2014).

45. Wang, P., Zhang, J., Shao, Q. & Wang, G. Physicochemical properties evolution of chars from palm kernel shell pyrolysis. *J. Therm. Anal. Calorim.* **133**, 1271–1280 (2018).

46. Pusceddu, E. *et al.* Chemical-physical analysis and exfoliation of biochar-carbon matter: From agriculture soil improver to starting material for advanced nanotechnologies. *Mater. Res. Express* **6**, 115612 (2019).

47. Xiao, X. & Chen, B. A Direct Observation of the Fine Aromatic Clusters and Molecular Structures of Biochars. *Environ. Sci. Technol.* **51**, 5473–5482 (2017).

48. Brown, R. A., Kercher, A. K., Nguyen, T. H., Nagle, D. C. & Ball, W. P. Production and characterization of synthetic wood chars for use as surrogates for natural sorbents. *Org. Geochem.* **37**, 321–333 (2006).

49. Zhao, S. X., Ta, N. & Wang, X. D. Effect of temperature on the structural and physicochemical properties of biochar with apple tree branches as feedstock material. *Energies* **10**, 1293 (2017).

50. Schneider, M. P. W., Hilf, M., Vogt, U. F. & Schmidt, M. W. I. The benzene polycarboxylic acid (BPCA) pattern of wood pyrolyzed between 200°C and 1000°C. *Org. Geochem.* **41**, 1082–1088 (2010).

51. Koch, B. P. & Dittmar, T. From mass to structure: an aromaticity index for high-resolution mass data of natural organic matter. *Rapid Commun. Mass Spectrom.* **20**, 926–932 (2006).

52. Xiao, X., Chen, Z. & Chen, B. H/C atomic ratio as a smart linkage between pyrolytic temperatures, aromatic clusters and sorption properties of biochars derived from diverse precursory materials. *Sci. Rep.* (2016) doi:10.1038/srep22644.

53. de Tomas, C., Suarez-Martinez, I. & Marks, N. A. Graphitization of amorphous carbons : A comparative study of interatomic potentials. *Carbon N. Y.* **109**, 681–693 (2016).

54. Franklin, R. E. Crystallite growth in graphitizing and non-graphitizing carbons. *Proc. R. Soc. London. Ser. A. Math. Phys. Sci.* **209**, 196–218 (1951).

55. Cohen-Ofri, I., Weiner, L., Boaretto, E., Mintz, G. & Weiner, S. Modern and fossil charcoal: Aspects of structure and diagenesis. *J. Archaeol. Sci.* **33**, 428–439 (2006).

56. McBeath, A. V. & Smernik, R. J. Variation in the degree of aromatic condensation of chars. *Org. Geochem.* **40**, 1161–1168 (2009).

57. Smernik, R. J., Kookana, R. S. & Skjemstad, J. O. NMR Characterization of 13 C-Benzene Sorbed to Natural and Prepared Charcoals. *Environ. Sci. Technol.* **40**, 1764–1769 (2006).

58. McBeath, A. V., Smernik, R. J., Krull, E. S. & Lehmann, J. The influence of feedstock and production temperature on biochar carbon chemistry: A solid-state 13C NMR study. *Biomass and Bioenergy* **60**, 121–129 (2014).

59. Wiedemeier, D. B. *et al.* Aromaticity and degree of aromatic condensation of char. *Org. Geochem.* (2015) doi:10.1016/j.orggeochem.2014.10.002.

60. McBeath, A. V., Smernik, R. J., Schneider, M. P. W., Schmidt, M. W. I. & Plant, E. L. Determination of the aromaticity and the degree of aromatic condensation of a thermosequence of wood charcoal using NMR. *Org. Geochem.* **42**, 1194–1202 (2011).

61. Kaal, J., Brodowski, S., Baldock, J. A., Nierop, K. G. J. & Cortizas, A. M. Characterisation of aged black carbon using pyrolysis-GC/MS, thermally assisted hydrolysis and methylation (THM), direct and cross-polarisation 13C nuclear magnetic resonance (DP/CP NMR) and the benzenepolycarboxylic acid (BPCA) method. *Org. Geochem.* **39**, 1415–1426 (2008).

62. Glaser, B., Haumaier, L., Guggenberger, G. & Zech, W. Black carbon in soils: The use of benzenecarboxylic acids as specific markers. *Org. Geochem.* **29**, 811–819 (1998).

63. Brodowski, S., Rodionov, A., Haumaier, L., Glaser, B. & Amelung, W. Revised black carbon assessment using benzene polycarboxylic acids. *Org. Geochem.* **36**, 1299–1310 (2005).

64. Vidal, J. L. *et al.* Green Solvents for the Liquid-Phase Exfoliation of Biochars. *ACS Sustain. Chem. Eng.* **9**, 9114–9125 (2021).

65. Cole, D. P., Smith, E. A. & Lee, Y. J. High-resolution mass spectrometric characterization of molecules on biochar from pyrolysis and gasification of switchgrass. in *Energy and Fuels* vol. 26 3803–3809 (American Chemical Society, 2012).

66. Martin, J. W. *et al.* Nanostructure of Gasification Charcoal (Biochar). (2019) doi:10.1021/acs.est.8b06861.

67. Zhao, N., Lv, Y. & Yang, X. A new 3D conceptual structure modeling of biochars by molecular mechanic and molecular dynamic simulation. *J. Soils Sediments* **17**, 641–655 (2017).

68. McKenna, A. M. *et al.* Expanding the Analytical Window for Biochar Speciation: Molecular Comparison of Solvent Extraction and Water-Soluble Fractions of Biochar by FT-ICR Mass Spectrometry. *Anal. Chem.* **93**, 15365–15372 (2021).

69. Oberlin, A. Carbonization and graphitization. *Carbon N. Y.* **22**, 521–541 (1984).

70. Harris, P. J. F. Fullerene-like models for microporous carbon. *J. Mater. Sci.* **48**, 565–577 (2013).

71. Harris, P. J. F. Structure of non-graphitising carbons. *Int. Mater. Rev.* **42**, 206–218 (1997).

72. Harris, P. J. F., Burian, A. & Duber, S. High-resolution electron microscopy of a microporous carbon. *Philos. Mag. Lett.* **80**, 381–386 (2000).

73. Guo, J. *et al.* Topological Defects: Origin of Nanopores and Enhanced Adsorption Performance in Nanoporous Carbon. *Small* **8**, 3283–3288 (2012).

74. McDonald-Wharry, J. S., Manley-Harris, M. & Pickering, K. L. Reviewing, Combining, and Updating the Models for the Nanostructure of Non-Graphitizing Carbons Produced from Oxygen-Containing Precursors. *Energy and Fuels* vol. 30 7811–7826 (2016).

75. Martin, J. W., De Tomas, C., Suarez-Martinez, I., Kraft, M. & Marks, N. A. Topology of disordered 3D graphene networks. *Phys. Rev. Lett.* **123**, 116105 (2019).

76. Smith, M. A., Foley, H. C. & Lobo, R. F. A simple model describes the PDF of a non-graphitizing carbon. *Carbon N. Y.* **42**, 2041–2048 (2004).

77. McDonald-Wharry, J., Manley-Harris, M. & Pickering, K. A comparison of the charring and carbonisation of oxygen-rich precursors with the thermal reduction of graphene oxide. *Philos. Mag.* **95**, 4054–4077 (2015).

78. Sharma, S., Shyam Kumar, C. N., Korvink, J. G. & Kübel, C. Evolution of Glassy Carbon Microstructure: In Situ Transmission Electron Microscopy of the Pyrolysis Process. *Sci. Rep.* **8**, 16282 (2018).

79. Harris, P. J. F. & Tsang, S. C. High-resolution electron microscopy studies of non-graphitizing carbons. *Philos. Mag. A Phys. Condens. Matter, Struct. Defects Mech. Prop.* **76**, 667–677 (1997).

80. Harris, P. J. F. Non-Graphitizing Carbon: Its Structure and Formation from Organic Precursors. *Eurasian Chem. J.* **21**, 227 (2019).

81. Vallejos-Burgos, F. *et al.* On the structural and reactivity differences between biomass- and coal-derived chars. *Carbon N. Y.* **109**, 253–263 (2016).

82. Pulido-Novicio, L. *et al.* Adsorption capacities and related characteristics of wood charcoals carbonized using a one-step or two-step process. *J. Wood Sci.* **47**, 48–57 (2001).

83. Marriott, A. S. *et al.* Investigating the structure of biomass-derived non-graphitizing mesoporous carbons by electron energy loss spectroscopy in the transmission electron microscope and X-ray photoelectron spectroscopy. *Carbon N. Y.* **67**, 514–524 (2014).

84. Deldicque, D., Rouzaud, J. N. & Velde, B. A Raman - HRTEM study of the carbonization of wood: A new Raman-based paleothermometer dedicated to archaeometry. *Carbon N. Y.* **102**, 319–329 (2016).

85. Rouzaud, J. N., Deldicque, D., Charon, É. & Pageot, J. Carbons at the heart of questions

on energy and environment: A nanostructural approach. *Comptes Rendus - Geosci.* **347**, 124–133 (2015).

86. Koo-amornpattana, W. *et al.* Valorization of spent disposable wooden chopstick as the $CO_2$ adsorbent for a $CO_2/H_2$ mixed gas purification. *Sci. Reports 2022 121* **12**, 1–16 (2022).

87. Matthews, M. J., Pimenta, M. A., Dresselhaus, G., Dresselhaus, M. S. & Endo, M. Origin of dispersive effects of the Raman D band in carbon materials. *Phys. Rev. B - Condens. Matter Mater. Phys.* **59**, R6585–R6588 (1999).

88. Kelemen, S. R. & Fang, H. L. Maturity trends in Raman spectra from kerogen and coal. *Energy and Fuels* **15**, 653–658 (2001).

89. Schito, A., Romano, C., Corrado, S., Grigo, D. & Poe, B. Diagenetic thermal evolution of organic matter by Raman spectroscopy. *Org. Geochem.* **106**, 57–67 (2017).

90. Sadezky, A., Muckenhuber, H., Grothe, H., Niessner, R. & Pöschl, U. Raman microspectroscopy of soot and related carbonaceous materials: Spectral analysis and structural information. *Carbon N. Y.* **43**, 1731–1742 (2005).

91. Smith, M. W. *et al.* Structural analysis of char by Raman spectroscopy: Improving band assignments through computational calculations from first principles. *Carbon N. Y.* **100**, 678–692 (2016).

92. Mukome, F. N. D., Zhang, X., Silva, L. C. R., Six, J. & Parikh, S. J. Use of chemical and physical characteristics to investigate trends in biochar feedstocks. *J. Agric. Food Chem.* **61**, 2196–2204 (2013).

93. Ocampo-Perez, R. *et al.* Synthesis of biochar from chili seeds and its application to remove ibuprofen from water. Equilibrium and 3D modeling. *Sci. Total Environ.* **655**, 1397–1408 (2019).

94. TUINSTRA F & KOENIG JL. RAMAN SPECTRUM OF GRAPHITE. *J. Chem. Phys.* **53**, 1126–1130 (1970).

95. Wopenka, B. & Pasteris, J. D. Structural characterization of kerogens to granulite-facies graphite: applicability of Raman microprobe spectroscopy. *Am. Mineral.* **78**, 533–557 (1993).

96. Keiluweit, M., Nico, P. S., Johnson, M. & Kleber, M. Dynamic molecular structure of

plant biomass-derived black carbon (biochar). *Environ. Sci. Technol.* **44**, 1247–1253 (2010).

97. Cetin, E., Moghtaderi, B., Gupta, R. & Wall, T. F. Influence of pyrolysis conditions on the structure and gasification reactivity of biomass chars. *Fuel* **83**, 2139–2150 (2004).

98. Mohanty, P. *et al.* Evaluation of the physiochemical development of biochars obtained from pyrolysis of wheat straw, timothy grass and pinewood: Effects of heating rate. *J. Anal. Appl. Pyrolysis* **104**, 485–493 (2013).

99. Bragg, W. L. The Diffraction of Short Electromagnetic Waves by a Crystal. in *Proceedings of the Cambridge Philosophical Society* 43–57 (1913).

100. Thomas, J. M. The birth of X-ray crystallography. *Nat. 2012 4917423* **491**, 186–187 (2012).

101. Scherrer, P. Bestimmung der Größe und der inneren Struktur von Kolloidteilchen mittels Röntgenstrahlen. *Nachrichten von der Gesellschaft der Wissenschaften zu Göttingen, Math. Klasse* **1918**, 98–100.

102. Marsh, H. & Rodríguez-Reinoso, F. *Activated Carbon. Activated Carbon* (Elsevier Ltd, 2006). doi:10.1016/B978-0-08-044463-5.X5013-4.

103. Lim, D. J., Marks, N. A. & Rowles, M. R. Universal Scherrer equation for graphene fragments. *Carbon N. Y.* **162**, 475–480 (2020).

104. Aoki, Y. *et al.* The role of the 2D-to-3D transition in x-ray diffraction analysis of crystallite size. *J. Phys. Condens. Matter* **33**, 294002 (2021).

105. Ban, L. L., Crawford, D. & Marsh, H. Lattice-resolution electron microscopy in structural studies of non-graphitizing carbons from polyvinylidene chloride (PVDC). *J. Appl. Crystallogr.* **8**, 415–420 (1975).

106. Rennhofer, H. *et al.* Pore Development during the Carbonization Process of Lignin Microparticles Investigated by Small Angle X-ray Scattering. *Molecules* **26**, 2087 (2021).

107. Boehm, H. P. Some aspects of the surface chemistry of carbon blacks and other carbons. *Carbon N. Y.* **32**, 759–769 (1994).

108. Boehm, H.-P., Diehl, E., Heck, W. & Sappok, R. Surface Oxides of Carbon. *Angew. Chemie Int. Ed. English* **3**, 669–677 (1964).

109. Boehm, H. P. Surface oxides on carbon and their analysis: a critical assessment. *Carbon N. Y.* **40**, 145–149 (2002).

110. Askeland, M., Clarke, B. & Paz-Ferreiro, J. Comparative characterization of biochars produced at three selected pyrolysis temperatures from common woody and herbaceous waste streams. *PeerJ* **7**, e6784 (2019).

111. Chandra, S. & Bhattacharya, J. Influence of temperature and duration of pyrolysis on the property heterogeneity of rice straw biochar and optimization of pyrolysis conditions for its application in soils. *J. Clean. Prod.* **215**, 1123–1139 (2019).

112. Kirtania, K., Tanner, J., Kabir, K. B., Rajendran, S. & Bhattacharya, S. In situ synchrotron IR study relating temperature and heating rate to surface functional group changes in biomass. *Bioresour. Technol.* **151**, 36–42 (2014).

113. Li, B. *et al.* Changes in Biochar Functional Groups and Its Reactivity after Volatile-Char Interactions during Biomass Pyrolysis. *Energy and Fuels* **34**, 14291–14299 (2020).

114. Trivedi, N. S., Mandavgane, S. A. & Chaurasia, A. Characterization and valorization of biomass char: a comparison with biomass ash. *Environ. Sci. Pollut. Res.* **25**, 3458–3467 (2018).

115. Paul, D., Kasera, N., Kolar, P. & Hall, S. G. Physicochemical characterization data of pine-derived biochar and natural zeolite as precursors to catalysts. *Chem. Data Collect.* **30**, 100573 (2020).

116. Leng, L. *et al.* An overview of sulfur-functional groups in biochar from pyrolysis of biomass. *Journal of Environmental Chemical Engineering* vol. 10 107185 (2022).

117. Perry, D. L. & Grint, A. Application of XPS to coal characterization. *Fuel* **62**, 1024–1033 (1983).

118. Suliman, W. *et al.* Modification of biochar surface by air oxidation: Role of pyrolysis temperature. *Biomass and Bioenergy* **85**, 1–11 (2016).

119. Leng, L. *et al.* Nitrogen containing functional groups of biochar: An overview. *Bioresource Technology* vol. 298 122286 (2020).

120. Suliman, W. *et al.* The role of biochar porosity and surface functionality in augmenting hydrologic properties of a sandy soil. *Sci. Total Environ.* **574**, 139–147 (2017).

121. Li, N. *et al.* Maximizing the number of oxygen-containing functional groups on activated

carbon by using ammonium persulfate and improving the temperature-programmed desorption characterization of carbon surface chemistry. *Carbon N. Y.* **49**, 5002–5013 (2011).

122. de la Puente, G., Pis, J. J., Menéndez, J. A. & Grange, P. Thermal stability of oxygenated functions in activated carbons. *J. Anal. Appl. Pyrolysis* **43**, 125–138 (1997).

123. Otake, Y. & Jenkins, R. G. Characterization of oxygen-containing surface complexes created on a microporous carbon by air and nitric acid treatment. *Carbon N. Y.* **31**, 109–121 (1993).

124. Zhuang, Q. L., Kyotani, T. & Tomita, A. DRIFT and TK/TPD Analyses of Surface Oxygen Complexes Formed during Carbon Gasification. *Energy and Fuels* **8**, 714–718 (1994).

125. Zielke, U., Hüttinger, K. J. & Hoffman, W. P. Surface-oxidized carbon fibers: I. Surface structure and chemistry. *Carbon N. Y.* **34**, 983–998 (1996).

126. Marchon, B., Carrazza, J., Heinemann, H. & Somorjai, G. A. TPD and XPS studies of O2, CO2, and H2O adsorption on clean polycrystalline graphite. *Carbon N. Y.* **26**, 507–514 (1988).

127. Figueiredo, J. L., Pereira, M. F. R., Freitas, M. M. A. & Órfão, J. J. M. Modification of the surface chemistry of activated carbons. *Carbon N. Y.* **37**, 1379–1389 (1999).

128. Samant, P. V., Gonçalves, F., Freitas, M. M. A., Pereira, M. F. R. & Figueiredo, J. L. Surface activation of a polymer based carbon. in *Carbon* vol. 42 1321–1325 (Pergamon, 2004).

129. Figueiredo, J. L. & Pereira, M. F. R. The role of surface chemistry in catalysis with carbons. *Catal. Today* **150**, 2–7 (2010).

130. Fan, M. *et al.* In situ characterization of functional groups of biochar in pyrolysis of cellulose. *Sci. Total Environ.* **799**, 149354 (2021).

131. Reeves, J. B. Mid-Infrared Spectroscopy of Biochars and Spectral Similarities to Coal and Kerogens: What Are the Implications? *Appl. Spectrosc.* **66**, 689–695 (2012).

132. Sbizzaro, M. *et al.* Effect of production temperature in biochar properties from bamboo culm and its influences on atrazine adsorption from aqueous systems. *J. Mol. Liq.* **343**, 117667 (2021).

133. Chen, Z., Xiao, X., Chen, B. & Zhu, L. Quantification of Chemical States, Dissociation Constants and Contents of Oxygen-containing Groups on the Surface of Biochars Produced at Different Temperatures. (2014) doi:10.1021/es5043468.

134. Li, Z. Q., Lu, C. J., Xia, Z. P., Zhou, Y. & Luo, Z. X-ray diffraction patterns of graphite and turbostratic carbon. *Carbon N. Y.* **45**, 1686–1695 (2007).

135. Felts, J. R. *et al.* Direct mechanochemical cleavage of functional groups from graphene. *Nat. Commun. 2015 61* **6**, 1–7 (2015).

136. Chang, D. W. *et al.* Solvent-free mechanochemical reduction of graphene oxide. *Carbon N. Y.* **77**, 501–507 (2014).

137. Burk, L. *et al.* Mechanochemical Routes to Functionalized Graphene Nanofillers Tuned for Lightweight Carbon/Hydrocarbon Composites. *Macromol. Mater. Eng.* **304**, 1800496 (2019).

138. Usman, A. R. A. *et al.* Biochar production from date palm waste: Charring temperature induced changes in composition and surface chemistry. *J. Anal. Appl. Pyrolysis* **115**, 392–400 (2015).

139. Baltrenas, P., Baltrenaite, E. & Spudulis, E. Biochar from pine and birch morphology and pore structure change by treatment in biofilter. *Water. Air. Soil Pollut.* **226**, 1–14 (2015).

140. De Boer, J. H. & Steggerda, J. . The helium density of micro-porous solids. in *Proceedings of the Koninklijke Nederlandse Akademie van Wetenschappen. Series B, Palaeontology, geology, physics and chemistry* 318–323 (Koninklijke Nederlandse akademie van wetenschappen, 1958).

141. Kini, K. A. & Stacy, W. O. The adsorption of helium by carbonaceous solids. *Carbon N. Y.* **1**, 17–24 (1963).

142. Menon, P. G. Adsorption at high pressures. *Chem. Rev.* **68**, 277–294 (1968).

143. Greyson, J. & Aston, J. G. The heats of adsorption of helium and neon on graphitized carbon black. *J. Phys. Chem.* **61**, 610–613 (1957).

144. Maggs, F. A. P., Schwabe, P. H. & Williams, J. H. Adsorption of Helium on Carbons: Influence on Measurement of Density. *Nat. 1960 1864729* **186**, 956–958 (1960).

145. Nguyen, H. G. T., Horn, J. C., Bleakney, M., Siderius, D. W. & Espinal, L.

Understanding material characteristics through signature traits from helium pycnometry. *Langmuir* **35**, 2115–2122 (2019).

146. Somerville, M. & Jahanshahi, S. The effect of temperature and compression during pyrolysis on the density of charcoal made from Australian eucalypt wood. *Renew. Energy* **80**, 471–478 (2015).

147. Giesche, H. Mercury Porosimetry: A General (Practical) Overview. *Part. Part. Syst. Charact.* **23**, 9–19 (2006).

148. Thommes, M. *et al.* Physisorption of gases, with special reference to the evaluation of surface area and pore size distribution (IUPAC Technical Report). *Pure Appl. Chem.* **87**, 1051–1069 (2015).

149. Maziarka, P. *et al.* Do you BET on routine? The reliability of N2 physisorption for the quantitative assessment of biochar's surface area. *Chem. Eng. J.* **418**, 129234 (2021).

150. Roussel, T., Jagiello, J., Pellenq, R. J. M., Thommes, M. & Bichara, C. Testing the feasibility of using the density functional theory route for pore size distribution calculations of ordered microporous carbons. *Mol. Simul.* **32**, 551–555 (2006).

151. Sigmund, G., Hüffer, T., Hofmann, T. & Kah, M. Biochar total surface area and total pore volume determined by N2 and CO2 physisorption are strongly influenced by degassing temperature. *Sci. Total Environ.* **580**, 770–775 (2017).

152. Cheng, D. *et al.* Feasibility study on a new pomelo peel derived biochar for tetracycline antibiotics removal in swine wastewater. *Sci. Total Environ.* **720**, 137662 (2020).

153. Brewer, C. E., Unger, R., Schmidt-Rohr, K. & Brown, R. C. Criteria to Select Biochars for Field Studies based on Biochar Chemical Properties. *Bioenergy Res.* **4**, 312–323 (2011).

154. Schlumberger, C. *et al.* Reliable surface area determination of powders and meso/macroporous materials: Small-angle X-ray scattering and gas physisorption. *Microporous Mesoporous Mater.* **329**, 111554 (2021).

155. Rouquerol, J., Llewellyn, P. & Rouquerol, F. Characterization of Porous Solids VII. **160**, 49–56 (2007).

156. Sigmund, G., Sun, H., Hofmann, T. & Kah, M. Predicting the Sorption of Aromatic Acids to Noncarbonized and Carbonized Sorbents. *Environ. Sci. Technol.* **50**, 3641–3648

(2016).

157. Břendová, K. *et al.* Biochar physicochemical parameters as a result of feedstock material and pyrolysis temperature: predictable for the fate of biochar in soil? *Environ. Geochem. Health* **39**, 1381–1395 (2017).

158. Kupgan, G., Liyana-Arachchi, T. P. & Colina, C. M. NLDFT Pore Size Distribution in Amorphous Microporous Materials. *Langmuir* **33**, 11138–11145 (2017).

159. Jagiello, J. *et al.* Exploiting the adsorption of simple gases O2 and H2 with minimal quadrupole moments for the dual gas characterization of nanoporous carbons using 2D-NLDFT models. *Carbon N. Y.* **160**, 164–175 (2020).

160. Villarroel-Rocha, J., Barrera, D., Arroyo-Gómez, J. J. & Sapag, K. Insights of adsorption isotherms with different gases at 77 K and their use to assess the BET area of nanoporous silica materials. *Adsorption* **27**, 1081–1093 (2021).

161. Blankenship, L. S., Jagiello, J. & Mokaya, R. Confirmation of pore formation mechanisms in biochars and activated carbons by dual isotherm analysis. *Mater. Adv.* **3**, 3961–3971 (2022).

162. UC Davis Biochar Database Home.

163. UK Biochar Research Centre. UKBRC Charchive.

164. Ippolito, J. A. *et al.* Feedstock choice, pyrolysis temperature and type influence biochar characteristics: a comprehensive meta-data analysis review. *Biochar* vol. 2 421–438 (2020).

165. Wang, L. *et al.* New trends in biochar pyrolysis and modification strategies: feedstock, pyrolysis conditions, sustainability concerns and implications for soil amendment. *Soil Use and Management* vol. 36 358–386 (2020).